\documentclass[prb,aps,twocolumn]{revtex4}
\usepackage{amsmath,amssymb,graphicx,subfigure}

\newcommand{\bea}{\begin{eqnarray}}
\newcommand{\eea}{\end{eqnarray}}
\newcommand{\be}{\begin{equation}}
\newcommand{\ee}{\end{equation}}
\newcommand{\nn}{\nonumber}

\begin{document}

\title{Screening of charged impurities with multi-electron singlet-triplet spin qubits in quantum dots}
\author{Edwin Barnes, J.~P.~Kestner, N.~T.~T.~Nguyen, and S.~Das Sarma}
\affiliation{Condensed Matter Theory Center, Department of Physics, University of Maryland, College Park, MD 20742}
\date{\today}
\begin{abstract}
Charged impurities in semiconductor quantum dots comprise one of the main obstacles to achieving scalable fabrication and manipulation of singlet-triplet spin qubits.  We theoretically show that using dots that contain several electrons each can help to overcome this problem through the screening of the rough and noisy impurity potential by the excess electrons.  We demonstrate how the desired screening properties turn on as the number of electrons is increased, and we characterize the properties of a double quantum dot singlet-triplet qubit for small odd numbers of electrons per dot.  We show that the sensitivity of the multi-electron qubit to charge noise may be an order of magnitude smaller than that of the two-electron qubit.
\end{abstract}
\maketitle
One of the most promising paths to scalable quantum computation is to use laterally defined double quantum dots (DQDs) in semiconductor heterostructures.  The qubit is formed by the spin states of the two-electron DQD with total spin projection zero along the $z$ axis.\cite{Levy02,Taylor05} Such a qubit is insensitive to spatially uniform magnetic field fluctuations, and, most importantly, amenable to fast electrical manipulation.\cite{Taylor05,Petta05} Recent experiments have made tremendous advances along these lines, demonstrating single-qubit initialization, arbitrary manipulation, and single-shot readout, all within a fraction of the coherence time of the qubit.\cite{Petta05,Barthel09,Foletti09,Barthel10,Bluhm11} Preliminary steps toward an entangling two-qubit gate have also been reported.\cite{Weperen11} In principle, successful completion of that program leaves the (admittedly enormous) challenge of scaling up to large numbers of qubits as the last remaining hurdle in the fabrication of a practical quantum computer.

However, a practical issue has emerged which threatens to be a crippling impediment to continued rapid progress.  The semiconductor samples used to create quantum dots invariably contain a number of charge impurity centers, perhaps $10^{10}\text{cm}^{-2}$ in GaAs systems.\cite{Nguyen11}  Even if the charge on these centers can be frozen to avoid switching noise, their presence inhibits access to the one-electron-per-dot regime since the lowest energy states of the dot may be fragmented due to the roughened potential landscape.\cite{Nixon90,Nagamune94,Chan04,LairdThesis}  This makes it difficult to find samples suitable for spin qubit realization.  Furthermore, typically the impurities do introduce some switching noise,\cite{Jung04,Pioro-Ladriere05,Buizert08,Petersson10} so that even in samples in which the impurities are all far enough from the DQD that a two-electron singlet-triplet qubit can be accessed, the interdot exchange energy is still subject to random fluctuation, leading to gate errors and decoherence.\cite{Hu06,Culcer09,Ramon10}  This necessitates operating in a parameter regime such that the sensitivity of the exchange energy to the charge noise is minimized, a so-called ``sweet spot".\cite{Stopa08}  In general, the charge noise problem is even more pernicious when performing two-qubit operations directly mediated by the Coulomb interaction, and one must again seek a sweet spot.\cite{Nielsen11,Yang11}  However, in practice, this strategy may not be sufficient since one typically cannot optimize over all noise channels simultaneously.\cite{Nielsen10}

An alternative approach to overcoming the charge noise problem is to define qubits with several electrons per dot, such that the ``core" electrons (electrons paired up into singlets and filling up the lowest single-particle states comprising the multi-electron ground state) serve to screen out the charge impurities, while each dot contributes the spin of a lone ``valence" electron to form the qubit.  Experimental studies of spin blockade\cite{Pioro-Ladriere03,Liu05} and excitation spectra\cite{Hatano08} have already been performed in multi-electron DQDs.  On the theoretical side, early work analyzed some of the subtleties involved with qubit manipulation in the context of multi-electron dots,\cite{Hu01,Vorojtsov04} highlighting, for example, the need for an external magnetic field in the case of singlet-triplet qubits in order to ensure that the qubit is well defined. We shall see that the use of multi-electron dots for spin-based quantum computation merits further consideration because of the favorable behavior of such qubits in the presence of charge impurities.

In this work, we demonstrate the effectiveness of multiple electrons per dot in screening a single charged impurity. We do this by performing a detailed numerical analysis employing the configuration interaction method with up to ten electrons in the single dot case and up to 14 electrons in the case of DQDs. In the case of a single quantum dot, we compute the ground state energy twice: once for a multi-electron state which takes into account interactions between the electrons and the impurity, and a second time for a multi-electron state which ignores these interactions. This yields two ground state energies whose difference constitutes a clear and convenient measure of the screening mechanism since it is directly related to the re-adjustment of the multi-electron wavefunction in response to the impurity. We find that the effect of the screening grows quickly as the number of electrons in the dot is increased.

In the case of DQDs, we calculate the exchange energy of a singlet-triplet qubit with three to seven electrons per dot and analyze its sensitivity to the charged impurity. To properly compare exchange energies obtained for different numbers of electrons, we first define a valence electron tunneling rate, and as we change the number of electrons, we tune the DQD potential to keep this tunneling rate invariant. This helps to isolate true multi-electron physics from other effects which occur as a byproduct of adding electrons to the system. We find that as the number of electrons is increased, the exchange energy becomes significantly less sensitive to the impurity. With five electrons per dot, for example, the sensitivity can be reduced by nearly an order of magnitude relative to the one-electron-per-dot case. Having five electrons per dot also appears substantially better than having three; this is due at least in part to the fact that, in a magnetic field (though not the high fields considered in Ref.~\onlinecite{Abolfath06}), there occurs some anomalous behavior arising from the spatial dependence of the phase of the relevant valence orbital.\cite{Lei10}

In contrast to previous theoretical studies of multi-electron DQDs\cite{Wensauer00,Szafran05,Li09} or multi-electron qubits,\cite{Hu01,Vorojtsov04,Scarola05} our focus is on how charge noise affects singlet-triplet qubits in a DQD. However, our results also give some new insights pertaining to the multi-electron quantum dot system in the absence of charge impurities, and so has some direct bearing on these earlier works. In particular, some of these works\cite{Hu01,Vorojtsov04} made use of the so-called ``frozen-core" approximation, which is an approximation often employed in the context of numerical solutions (particularly the configuration interaction method) to multi-electron quantum dot problems. In this approximation, one keeps the core electrons frozen in the lowest non-interacting states of the dot(s) and only allows configurations where the valence electrons are free to occupy higher energy levels. In the course of our analysis of the charged impurity screening effect, we will implement a more sophisticated approximation scheme wherein excitations of the core electrons are also taken into account, allowing us to test the accuracy of the frozen-core approximation. We find that the frozen-core approximation becomes significantly worse as the number of electrons is increased.  Although the method we employ constitutes a marked improvement over previous approaches, we focus on qualitative results and general trends since the microscopic details of the actual experimental potential and noise are not precisely known.

The structure of the paper is as follows:  In Sec.~\ref{sec:methods}, we explain our numerical methods and examine the validity of the approximations employed.  In Sec.~\ref{sec:screening}, we show the onset of screening effects in a multi-electron single quantum dot.  In Sec.~\ref{sec:stqubits}, we characterize multi-electron singlet-triplet qubits in DQDs and their sensitivity to charge noise.  Finally, we conclude in Sec.~\ref{sec:conclusions}. In Appendix \ref{app:A}, we review Fock-Darwin states to keep the paper self-contained and to fix conventions. Appendices \ref{app:B} and \ref{app:C} contain technical details that facilitate the numerical computations of the multi-electron spectra.
\section{Numerical methods}\label{sec:methods}
\subsection{Basis states}\label{subsec:basis}
All of our results are obtained in a configuration-interaction (CI) approach using molecular orbitals.  The Hamiltonian for $N$ electrons in confining potential $V\left(\mathbf{r} \right)$, in the presence of $M$ static charge impurities located at positions $\mathbf{R}_j$, and with a perpendicular magnetic field $\mathbf{B} \!=\! B \mathbf{\hat{z}}\!=\! \mathbf{\nabla \times A}$, is
\begin{multline}\label{eq:H}
H = \sum_{i=1}^N \left[ \frac{\left(-i\hbar\mathbf{\nabla_i} + e\mathbf{A}/c\right)^2}{2m^{\ast}} + V\left(\mathbf{r_i} \right)
+ \sum_{j=1}^M \frac{Z_{\text{imp}}e^2}{\epsilon |\mathbf{r_i}-\mathbf{R_j}|} \right]
\\
+ \sum_{i< j} \frac{e^2}{\epsilon |\mathbf{r_i}-\mathbf{r_j}|} + g^{\ast}\mu_B \mathbf{B\cdot S},
\end{multline}
where $m^{\ast}$ is the effective mass of the electrons, $Z_{\text{imp}}$ is the impurity strength, $\epsilon$ is the dielectric constant of the semiconductor, $g^{\ast}$ is the effective electron $g$-factor, $\mu_B$ is the Bohr magneton, and $\mathbf{S}$ is the total electronic spin.  In this work we will take parameters relevant to GaAs systems: $m^{\ast} \!=\! 0.067 m_e$, $g^{\ast}\!=\!-0.44$, $\epsilon \!=\! 13.1$.  We neglect perturbative corrections to the Hamiltonian such as spin-orbit coupling.  When including the effects of an impurity, we will consistently take $Z_{\text{imp}}\!=\!1$.

\begin{figure}
  \includegraphics[width=.8\columnwidth]{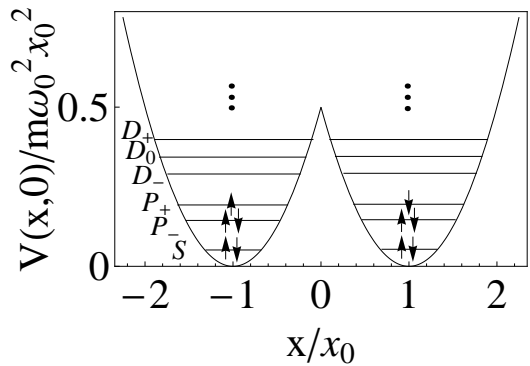}
  \caption{Model double well potential along the interdot axis.  There is also a harmonic potential along $y$ with frequency $\omega_0$.  We have schematically shown 10 electrons filling the lowest Fock-Darwin orbitals in a finite magnetic field.}\label{fig:potential}
\end{figure}
We will assume tight confinement along the $z$-direction in the depletion layer of a heterostructure interface and consider the resultant two-dimensional electron gas.  For simplicity, we will use a (bi-) quadratic in-plane potential to model the lateral gate-defined confinement of a (double) quantum dot,
\be\label{eq:potential}
V\left(x,y\right) = \frac{1}{2}m^{\ast}\omega_0^2 \text{Min}\{\left(x-x_0\right)^2+y^2,\left(x+x_0\right)^2+y^2\}
\ee
as shown in Fig.~\ref{fig:potential}.  We have checked that, as expected, the discontinuity in the derivative of the potential does not have any qualitative effect.  Although the actual confinement potential depends on the experimental details of the gate geometry and voltage, the biquadratic approximation has been justified by comparing to an exact numerical solution of Poisson's equation for a realistic setup.\cite{Nielsen10}  The appropriate single-particle orbitals are then the Fock-Darwin states centered at the minima of each well, $\mathbf{r}\!=\!\mp x_0 \hat{\mathbf{x}}$, labeled with a $\pm$ sign, respectively, and carrying principal quantum number $n\!=\!0,1,2,...$ and magnetic quantum number $m\!=\!-n,-n+2,...,n-2,n$,
\begin{multline}\label{eq:phi}
\phi_{nm}^{\pm}\left(x,y\right) = \frac{1}{\ell_0}\sqrt{\frac{\left(\frac{n-|m|}{2}\right)!}{\pi\left(\frac{n+|m|}{2}\right)!}} \left(\frac{x\pm x_0 + i y \, \text{sgn}\, m}{\ell_0}\right)^{|m|}
\\
\times e^{-\frac{\left(x \pm x_0\right)^2 + y^2}{2 \ell_0^2} \pm i\frac{x_0 y}{2\ell_B^2}} L_{\frac{n-|m|}{2}}^{|m|}\left(\frac{\left(x \pm x_0\right)^2 + y^2}{\ell_0^2}\right),
\end{multline}
where $\ell_0 \!=\! \ell_B/\left(1/4+\omega_0^2/\omega_c^2\right)^{1/4}$, $\ell_B \!=\! \sqrt{\hbar c/eB}$, $\omega_c \!=\! eB/m^{\ast}c$, $L_n^{m}\left(x\right)$ is an associated Laguerre polynomial, and we have taken the symmetric gauge, $\mathbf{A} \!=\! B/2 \left(-y \mathbf{\hat{x}} + x \mathbf{\hat{y}}\right)$.  The associated single-particle energies are
\be\label{eq:E}
E_{n,m} = \left(n + 1\right)\sqrt{\frac{1}{4} + \frac{\omega_0^2}{\omega_c^2}} \, \hbar \omega_c + \frac{m}{2} \hbar \omega_c.
\ee
The derivation of the Fock-Darwin states and spectrum is reviewed in Appendix \ref{app:A}. In analogy with the hydrogen atom, we refer to the $(n,m)\!=\!(0,0)$ state as the ``$S$" orbital, the $(n,m)\!=\!(1,\pm 1)$ state as the ``$P_{\pm}$" orbital, and so on.  We will similarly refer to groups of states with the same $n$ as ``shells."

The basis states used to expand the many-electron wavefunction are Slater determinants formed from the single-particle orbitals,
\be\label{eq:psi}
\Psi_{\{n_i m_i\}}^{\{s_i\}}\left(\{x_i\},\{y_i\},\{\sigma_i\}\right) = \mathcal{P}\biggl\{\prod_{i=1}^N \phi_{n_i m_i}^{s_i}\left(x_i,y_i\right)\chi_i\left(\sigma_i\right) \biggr\},
\ee
where $\sigma$ is the electronic spin variable, $\chi\left(\sigma\right)$ is the associated spin wavefunction, and $\mathcal{P}$ is the antisymmetrization operator.  See the appendices for a complete presentation of the basis states and useful matrix elements between them.

\subsection{Truncation and convergence}\label{subsec:convergence}
While the exact ground state can be built from the complete set of configurations using all combinations of Fock-Darwin single-particle states, obviously in practice one must truncate the space of possible configurations.  The analogy with the hydrogen atom spectrum naturally leads one to the idea that perhaps, in dealing with multi-particle states, one can treat the core electrons as being effectively inert.  In other words, one can consider the ``frozen-core" approximation in which one keeps only multi-particle states for which all but two of the electrons fill up the Fock-Darwin states below the valence orbital.\cite{Hu01} Another technique is to simply neglect all configurations involving single-particle orbitals above some cutoff level.  This is referred to as ``full CI" within the orbital cutoff.  A more efficient method is to neglect all configurations whose non-interacting energy is higher than the non-interacting ground state by some cutoff energy.  We have used each of these approximations in our numerical calculations for comparison.  Particularly for several electrons and in the presence of an impurity, the frozen-core approximation may not allow enough flexibility in the multi-electron wavefunction to accurately represent the ground state.

In a single dot with typical radius on the order of $100$ nm,\cite{Barthel09,Foletti09,Petta05} the on-site Coulomb energy is much greater than the non-interacting level spacing, so using the basis built from non-interacting Fock-Darwin orbitals requires a very large number of excited configurations to be kept in the CI calculation.  Since the ratio of Coulomb energy to level spacing is proportional to dot size, we take a relatively tight confinement potential with oscillator length between $6-10$ nm to aid convergence.  Even for these strongly confined electrons, though, the interaction energy scale is comparable to the confinement energy scale and we must keep many excited configurations.  Experimental dot radii are usually considerably larger,\cite{Barthel09,Foletti09,Petta05} but we expect that the features will be qualitatively similar.

\begin{figure}
  \includegraphics[width=.8\columnwidth]{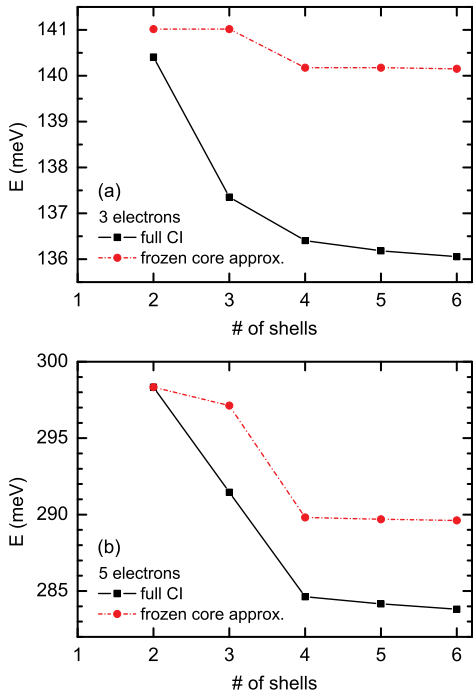}
  \caption{(Color online.) Ground state energy vs.~number of shells kept in the full CI calculation for three (a) and five (b) electrons in a harmonic trap with frequency $\hbar\omega_0 \!=\!24.26$ meV (a) and $\hbar\omega_0 \!=\!21.79$ meV (b) and perpendicular magnetic field $B\!=\!2.8$ T (a) and $B\!=\!2.5$T (b) such that $\omega_c\!=\!2\omega_0/\sqrt{99}$.}\label{fig:convergencefullCI}
\end{figure}

In Fig.~\ref{fig:convergencefullCI}, we show the convergence of the ground state energy for three and five electrons in the absence of impurities within a full CI calculation as the number of shells kept increases.  For five electrons, keeping six shells amounts to keeping nearly 20,000 configurations.  We have chosen values of the magnetic field such that $\omega_c \!=\! 2\omega_0/\sqrt{99}$, which, although it corresponds to a $\sim 3$T field for our unusually small dots, for a typical dot size corresponds to a field of only tens or hundreds of mT.  (The convergence results are specific to the choice of parameters.  The particular set of parameters taken here is in analogy to the DQD potential discussed in Sec.~\ref{subsec:params}.)  Also shown are the energies obtained using a frozen-core approximation in which only the outermost electron is free to occupy the excited shells.  Clearly the frozen-core results are significantly improved upon by allowing configurations with excited core electrons.  Again, this is due to the strong electron-electron interactions.

\begin{figure}
  \includegraphics[width=.8\columnwidth]{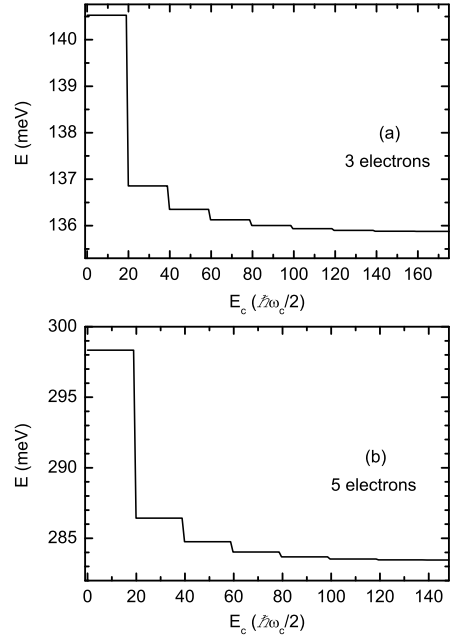}
  \caption{(Color online.) Ground state energy vs.~cutoff in the energetically truncated CI calculation for three (a) and five (b) electrons in a harmonic trap with parameters as in Fig.~\ref{fig:convergencefullCI}.}\label{fig:convergence}
\end{figure}

In Fig.~\ref{fig:convergence}, we keep all configurations with non-interacting energy within a given cutoff excitation energy, $E_c$, from the non-interacting ground state energy. For example, when $E_c\!=\!0$, only the ground-state configuration is kept. We show the convergence of the ground state energy versus cutoff for three and five electrons.  We use $\hbar\omega_c/2$ as a natural unit because that is the smallest increment by which the single-particle energies of Eq.~\eqref{eq:E} can change. The step-like behavior is due to the discreteness of the spectrum and the fact that angular momentum is a good quantum number: Increasing the cutoff energy only increases the basis size at a discrete set of points, and a new configuration with the correct angular momentum is added to the basis only when the cutoff permits configurations with an even increment in the total principal quantum number, $\sum_i n_i$.  For the parameters chosen, this occurs every $20 \hbar\omega_c/2$.  We see that taking an excitation cutoff of $E_c\!\sim\! 80\hbar\omega_c/2$ gives the same energy as a full CI approach with six shells, while only keeping about a tenth of the number of configurations.  However, even a cutoff of $E_c\!\sim\! 20\hbar\omega_c/2$ already captures the dominant corrections to the non-interacting state.  Somewhat similar convergence properties were noted in Ref.~\onlinecite{Reimann00}.

Similar calculations for the tunnel-coupled DQD system (with twice as many electrons) become computationally demanding at high cutoffs, but from the above results for a single dot (or equivalently, for well separated DQDs) we expect reasonable convergence is attained already for full CI up to the $F$ orbitals or for a cutoff of $E_c \!\sim\! 20\hbar\omega_c/2$.  Since full CI up to $F$ orbitals quickly becomes unwieldy for several electrons, we shall primarily use the cutoff approach for the DQD calculations.  For five electrons per dot, taking numerically tractable cutoffs on the order of $20\hbar\omega_c/2$ corresponds to keeping roughly the lowest thousand non-interacting DQD configurations, which we note to be a significant improvement on the common approximation of keeping only $S$, $P$, and sometimes $D$ shells with a frozen core. We will discuss DQD convergence further in Sec.~\ref{subsec:J}.

\section{Onset of screening for multi-electron dots}\label{sec:screening}
\begin{figure}
  \includegraphics[width=.8\columnwidth]{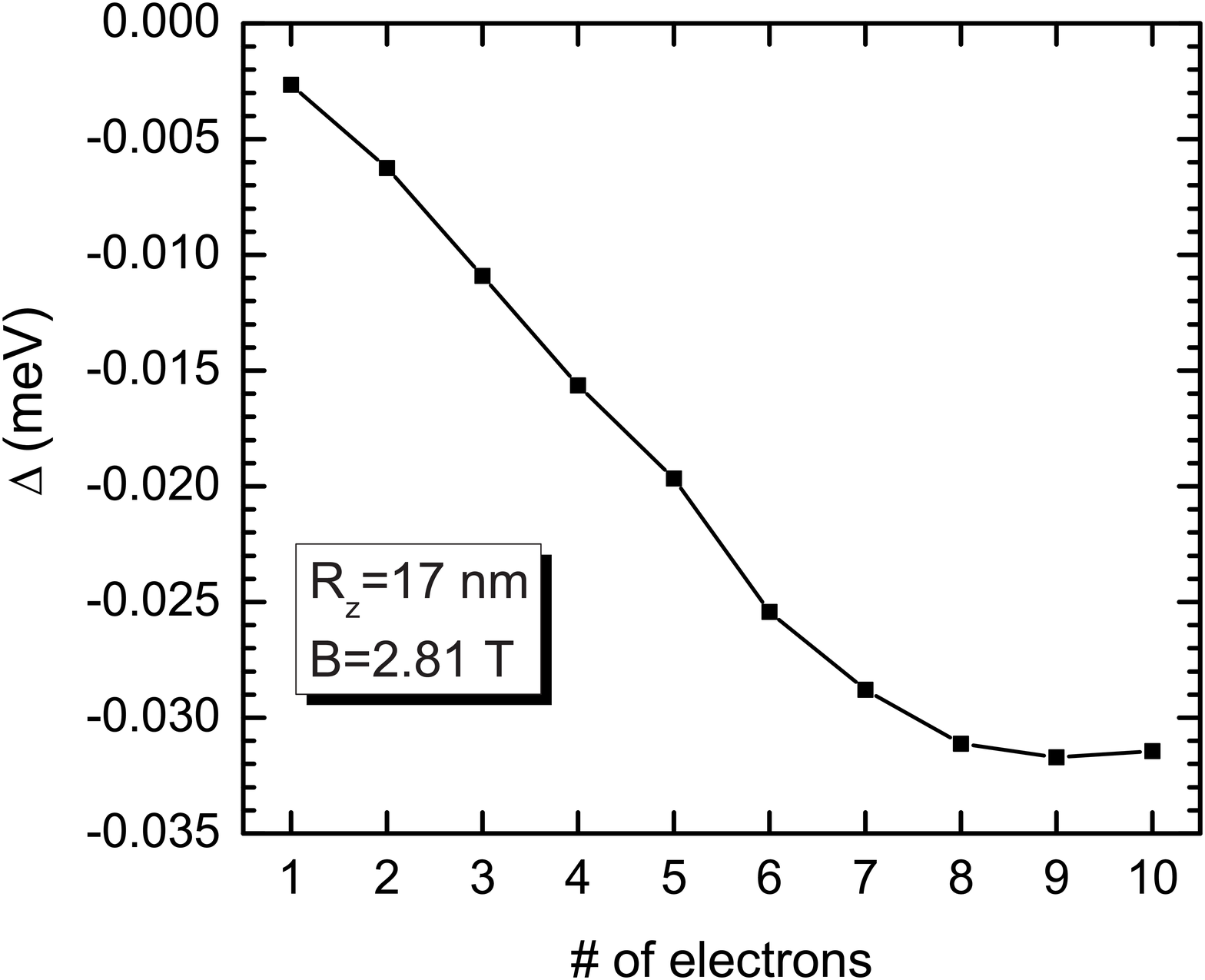}
  \includegraphics[width=.8\columnwidth]{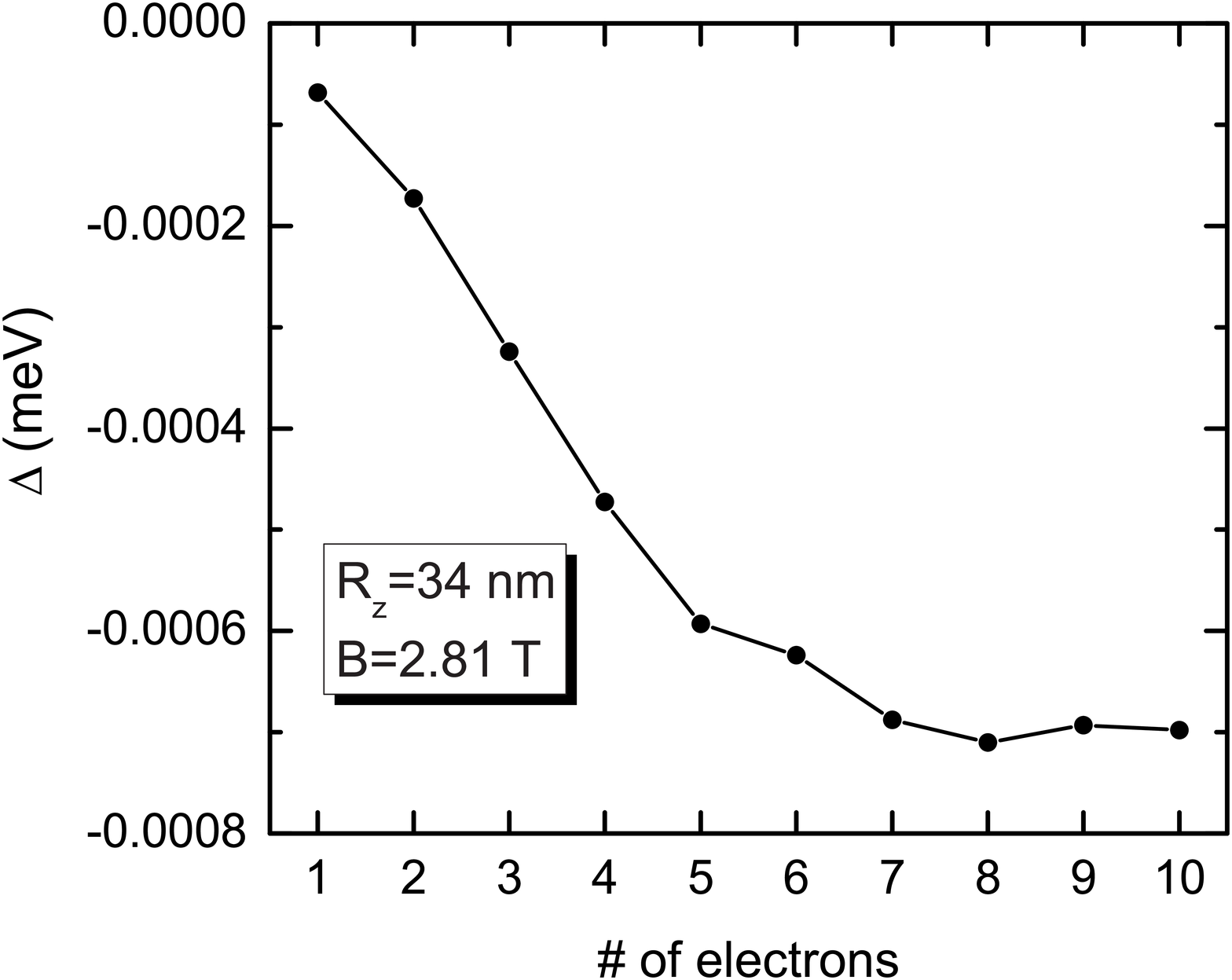}
  \caption{Rearrangement energy $\Delta$ (see text) vs.~number of electrons for an impurity on the $z$-axis.  Here the impurity is repulsive, and is located 17 nm (top) or 34 nm (bottom) from the dot center. There is a perpendicular field $B\!=\!2.8$T and the trap frequency is $\hbar\omega_0\!=\!24.26 meV$.}\label{fig:screeningz}
\end{figure}

\begin{figure}
  \includegraphics[width=.8\columnwidth]{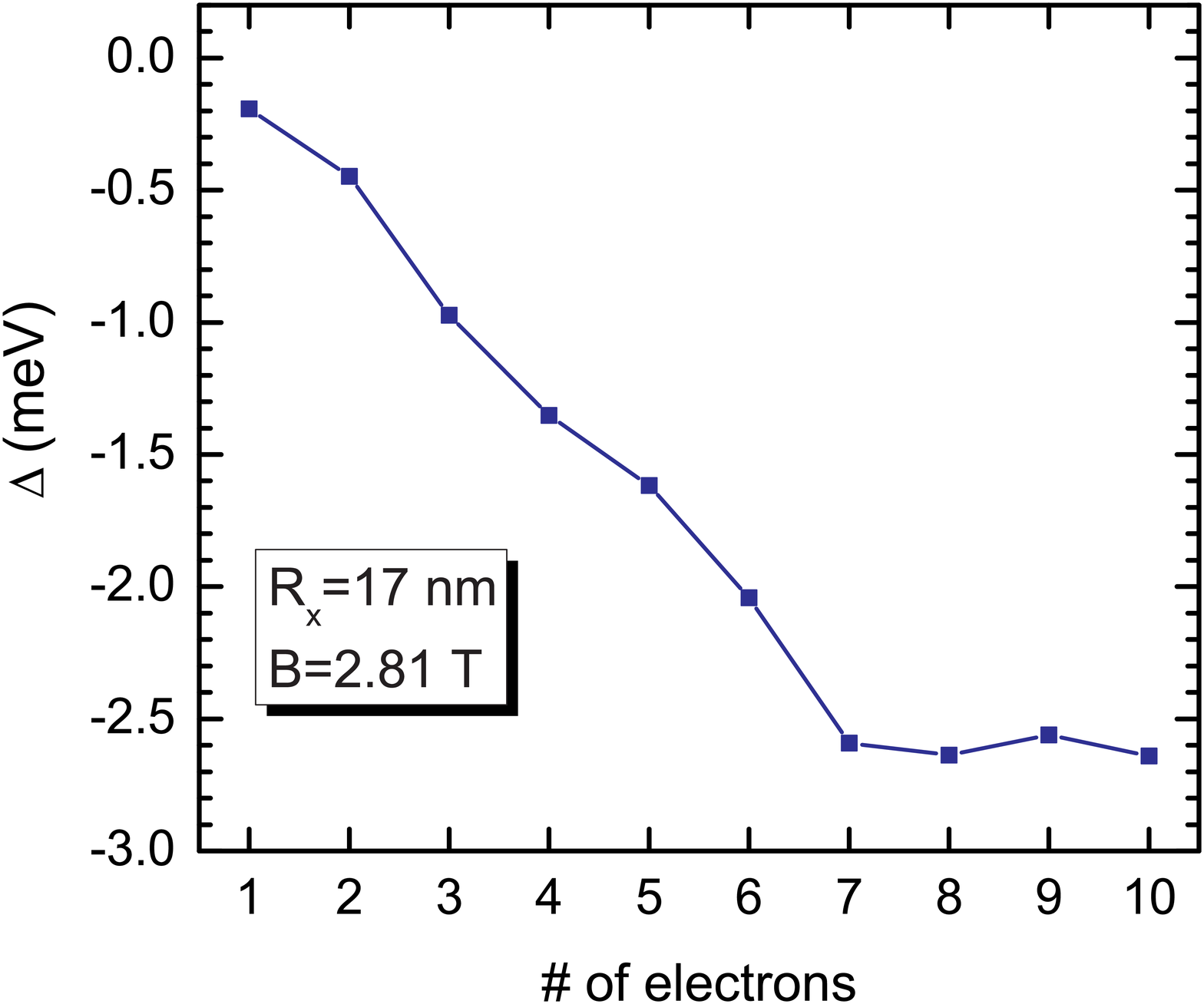}
  \includegraphics[width=.8\columnwidth]{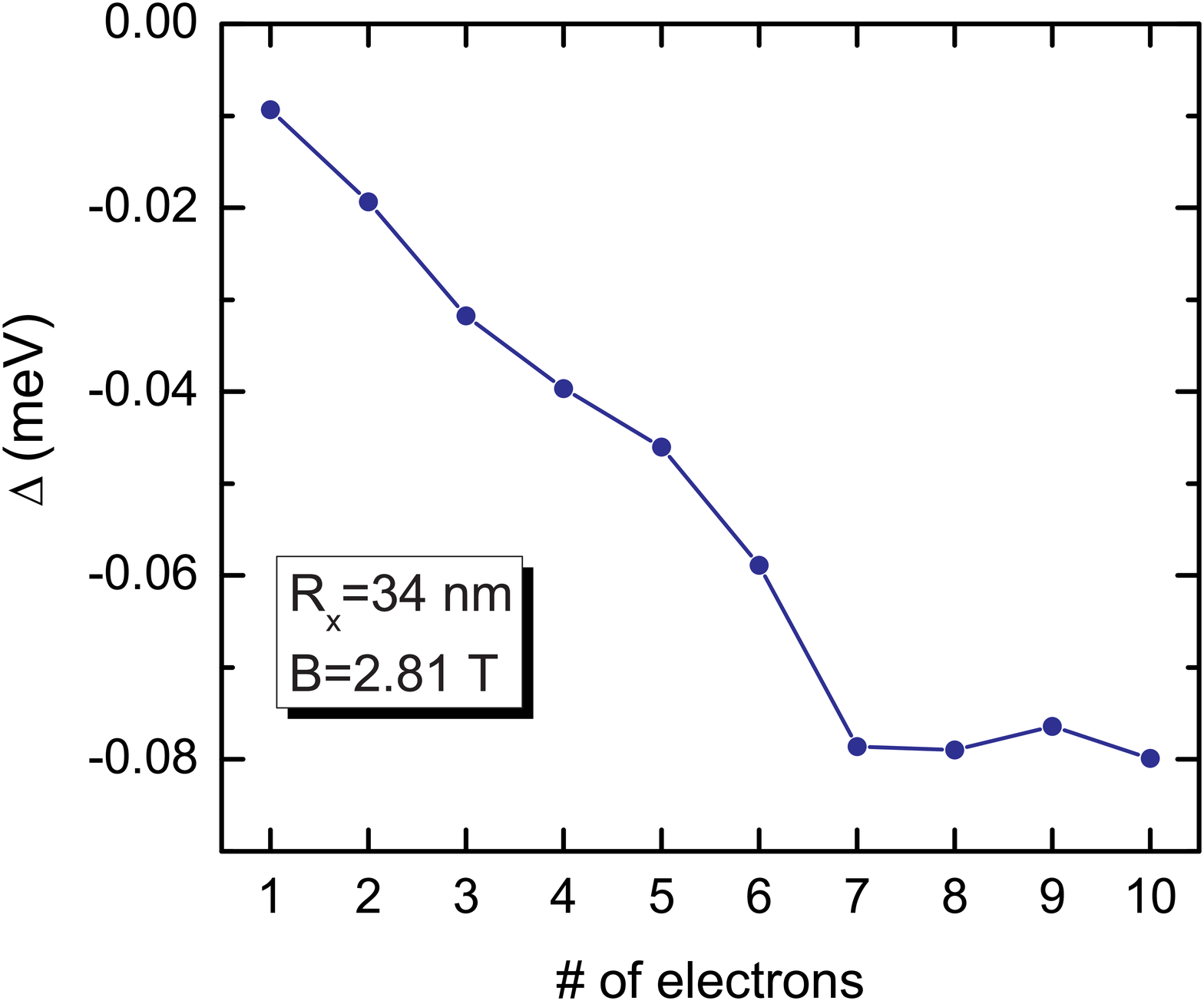}
  \caption{Rearrangement energy $\Delta$ (see text) vs.~number of electrons for an impurity on the $x$-axis.  Parameters are the same as in Fig.~\ref{fig:screeningz}.}\label{fig:screeningx}
\end{figure}

We now consider the multi-electron single quantum dot in the presence of a single charged impurity.  Upon turning on the impurity potential, the electrons redistribute themselves to minimize their total energy.  The amount by which the energy in the presence of the impurity changes due the rearrangement of the multi-electron wavefunction is
\be\label{eq:screening}
\Delta = \langle \Psi| H |\Psi \rangle - \langle \Psi_{0}| H |\Psi_{0} \rangle,
\ee
where $|\Psi\rangle$ and $|\Psi_{0}\rangle$ are the ground states with and without an impurity present, respectively, and $H$ is the Hamiltonian with an impurity.  This quantity is plotted in Figs.~\ref{fig:screeningz} and \ref{fig:screeningx} for different positions of the impurity.  The calculation is performed with cutoffs in the range $E_c \!=\! 45\hbar\omega_c/2$ to $E_c \!=\! 110\hbar\omega_c/2$, with lower cutoffs used for higher electron numbers.  For Fig.~\ref{fig:screeningz}, the minima of the combined impurity plus harmonic potential remains at the origin, and $\Delta$ may be thought of as a measure of the change in the shape of the wavefunction.  For Fig.~\ref{fig:screeningx}, the minima of the potential shifts, and $\Delta$ is much larger due to a trivial contribution coming from the translation of the wavefunction.

In both cases, the linear part of the curves may be understood as arising from the impurity pushing the electrons away from the center of the harmonic potential.  However, note that one begins to see marked deviations from linearity already at six or seven electrons\cite{note}.  This implies that nontrivial screening effects are already setting in whereby the first few electrons form a screening core which rearranges to partially shield any additional outer electrons from the impurity potential.  In other words, defining a multi-electron spin qubit using even a small number of electrons per dot is advantageous since screening reduces the effect of the impurity.  This is consistent with previous qualitative predictions of screening effects in the optical absorption of few-electron dots\cite{Halonen96} and in the high magnetic field limit of a six-electron dot with a repulsive impurity.\cite{Rasanen04}  We shall see in Sec.~\ref{subsec:dJ} that possible advantages are not limited to qubit fabrication in the presence of static impurities, but also apply to qubit operation in the presence of fluctuating impurities.

\section{Multi-electron singlet-triplet qubit in the presence of charge noise}\label{sec:stqubits}
The screening discussed in the previous section may aid the fabrication of multi-electron quantum dot systems in disordered background potential landscapes that would fragment or destroy single-electron dots.  For the purposes of quantum computation, we now wish to discuss the properties of a multi-electron singlet-triplet qubit that can take advantage of this screening.

\subsection{Qubit definition}\label{subsec:qubitdef}
Consider a DQD with an odd number of electrons in each dot.  In the absence of impurities and without the Coulomb interaction between electrons, the electrons forming the ground state will pair up into singlets and inhabit the lowest Fock-Darwin states.  If the tower of Fock-Darwin states in each dot is non-degenerate (which requires a non-zero magnetic field), then for odd numbers of electrons in each dot it will always be the case that there is exactly one highest occupied valence state in each dot, and the valence state in each dot contains only one ``valence" electron.  For $N\!=\!2,6,10,14$ electrons ($N$ is the total number of electrons in both dots), the valence state is the $S$ ($n\!=\!0$, $m\!=\!0$), $P_-$ ($n\!=\!1$, $m\!=\!-1$), $P_+$ ($n\!=\!1$, $m\!=\!1$), and $D_-$ ($n\!=\!2$, $m\!=\!-2$) orbital, respectively. The spins of the valence electrons can then be used to define a singlet-triplet qubit in exact analogy to the standard two-electron case.\cite{Taylor05} Even when the intuitive picture of an inert core breaks down, so that there is no sharp distinction between valence and core electrons, the qubit is still defined in terms of the lowest two many-body energy levels and the low-energy spectrum of the DQD retains a good qubit subspace as will be shown below.

\subsection{Parameter choices}\label{subsec:params}
As mentioned previously, for better convergence we take small dots with characteristic radius between $6-10$ nm. We choose the inter-dot distance to be $40$ nm in order to have distinct dots that still have appreciable tunnel coupling and exchange.  This is a smaller inter-dot distance than is typical for experiments,\cite{Petta05,Barthel09,Foletti09} but that is simply because the dots themselves are smaller here.  The qualitative behavior should be similar.

When comparing cases with different electron numbers, one should first isolate the intrinsic multi-electron effects from trivial filling effects.  For example, for a given double well potential, a six-electron singlet-triplet qubit will trivially have a larger exchange energy than a two-electron one simply because the valence electrons lie higher in the well and thus see a lower barrier to tunneling.  This hardly makes for an enlightening comparison between the two- and six-electron cases, since the dominant effect of naively adding more electrons is effectively nothing more than a change in the central barrier height.  We will instead compare different fillings on a more nuanced and consistent basis by adjusting the magnetic field and the confinement potential such that the tunneling between valence shells and the ratio of orbital level spacings remain constant as the filling is varied.

More precisely, we define the valence tunnel coupling for $N$ electrons in two dots to be the energy splitting in the frozen-core approximation between the two lowest states for $N\!-\!1$ interacting electrons in two dots, the idea being that this energy splitting characterizes the tunneling of an electron in the valence state of one dot to the valence state of the other.  This tunneling incorporates the Coulomb interactions with all core electrons, but does not include interactions with impurities.  By adjusting the confinement strength and the magnetic field, we fixed the value of this tunneling to be $0.2$ meV for all cases.  This particular value was chosen to achieve good numerical convergence while remaining in the window of experimentally relevant exchange energies, which are typically on the order of a few $\mu$eV;\cite{Petta05} in our setup, this corresponds to tunneling rates on the order of hundreds of $\mu$eV.

Before quoting the parameters needed to obtain this tunneling for the cases of $N\!=\!2, 6, 10, 14$ electrons, we first need to say a few words about the spectrum.  We mentioned above that we need an external magnetic field to obtain a non-degenerate spectrum in each dot.  However, we did not specify precisely what value this $B$-field should take.  We want the $B$-field to be large enough to avoid possible near-degeneracy effects, but at the same time, we do not want it too large because then different shells would start to overlap in energy.  Of course, the latter effect happens for any finite $B$-field at sufficiently large $n$.  In practice though, very high shells should not be relevant, and for the most part we will only keep states up through the $n\!=\!3$ shell. (For some of the $N=14$ results, we will find it necessary to also include some of the $n\!=\!4$ states, but this has little bearing on the present considerations and will be further clarified below in the context of the particular results in question.) It then suffices to make the $B$-field small enough so as to avoid any overlap between the $n\!=\!3$ ``$F$" shell and the $n\!=\!4$ ``$G$" shell.  We can satisfy all the above criteria by choosing the cyclotron frequency
\be
\omega_c = \frac{2}{\sqrt{99}} \omega_0 ,
\ee
leading to the Fock-Darwin spectrum of Table \ref{table:spectrum} and non-interacting multi-particle energies
\be\label{eq:Enonint}
E_{\{n_i\},\{m_i\}}^{\text{non-int}} = \sum_{i=1}^N E_{n_i,m_i} = \sum_{i=1}^N \left(10 n_i + 10 + m_i\right)\frac{\hbar\omega_c}{2}.
\ee
For this choice of parameters, the spacings between states of the same shell are smaller than the spacings between shells, but not drastically so.  Also notice that we are tying the $B$-field to the size of the wells.  This means that when we change the number of electrons, we will simultaneously change both the $B$-field and well size in such a way that the valence tunneling (as defined above) remains constant at $0.2$ meV.
\begin{table}
  \centering
  \begin{tabular}{c|c|c|c|c|c|c|c|c|c|c|c}
     & $S$ & $P_-$ & $P_+$ & $D_-$ & $D_0$ & $D_+$ & $F_{--}$ & $F_-$ & $F_+$ & $F_{++}$ & $G_{--}$ \\
    \hline
    $E/\left(\hbar\omega_c/2\right)$ & $10$ & $19$ & $21$ & $28$ & $30$ & $32$ & $37$ & $39$ & $41$ & $43$ & $46$ \\
  \end{tabular}
  \caption{Energies of low-lying Fock-Darwin shells for $\omega_c/\omega_0 = 2/\sqrt{99}$.}\label{table:spectrum}
\end{table}

\begin{figure}
  \includegraphics[width=.8\columnwidth]{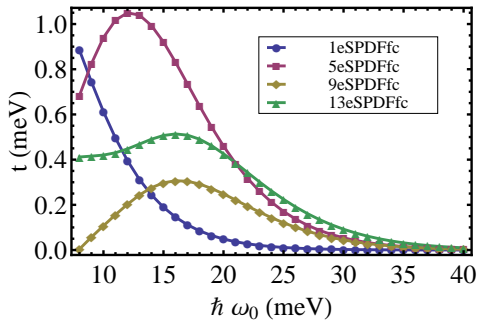}
  \caption{(Color online.) Tunnel coupling versus well depth for 1, 5, 9, and 13 electrons in two dots in the frozen-core approximation.}\label{fig:t}
\end{figure}

With these considerations, we calculated the frozen-core spectrum of $1$, $5$, $9$, and $13$ electrons in a double well with interdot spacing of $40$ nm to obtain the characteristic tunneling energies shown in Fig.~\ref{fig:t} as a function of confinement strength.  Note that the unphysical behavior of the tunneling for small $\omega_0$ is an artifact of the truncation of the configuration space, an approximation which clearly breaks down for weak confinement.  From Fig.~\ref{fig:t} we obtain the parameters (listed in Table \ref{table:params}) necessary to equalize the valence tunneling at $0.2$ meV for the $2$, $6$, $10$, and $14$-electron singlet-triplet qubits we wish to consider.

Note also from Fig.~\ref{fig:t} that the valence tunneling is actually smaller for nine electrons, where the valence orbital is $P_+$, than for five electrons, where the valence orbital is $P_-$.  It is rather unusual and counterintuitive that a higher energy orbital has a smaller tunneling rate!  We interpret this as a consequence of the phase $e^{\pm ix_0 y/(2\ell_B^2)}$ appearing in the $n\!=\!1$ shell hopping-type integrals, $\int d\mathbf{r} \left(\partial_x \phi_{1,\pm 1}^{+}\right)^\ast \partial_x \phi_{1,\pm 1}^{-}$, due to the magnetic field (see Eq.~\eqref{eq:phi}). This allows contributions from the prefactor in the integrand which is an odd function of $y$.  Since, for $x_0 > \ell_0$,  the odd term has a $+$ ($-$) sign for $m\!<\!0$ ($m\!>\!0$), the magnetic field diminishes the tunneling of the higher $P$ orbital and enhances that of the lower one. Somewhat similar considerations have been noted in recent experimental work.\cite{Lei10}

\begin{table}
  \centering
  \begin{tabular}{c|c|c}
    $N$ & $\hbar\omega_0$ (meV) & $B$ ($T$) \\
    \hline
    2 & 14.78 & 1.7 \\
    6 & 24.26 & 2.8 \\
    10 & 21.79 & 2.5 \\
    14 & 25.71 & 3.0 \\
  \end{tabular}
  \caption{Parameters for which the valence tunneling is $t\!=\!0.2$ meV with interdot distance fixed at $40$ nm.}\label{table:params}
\end{table}

\subsection{Qubit characterization}\label{subsec:J}
We characterize these multi-electron singlet-triplet qubits by calculating the energy splitting, $J$, between the two lowest levels in the absence of a charged impurity.  (These are well separated from the other levels, as shown below, and form a good qubit subspace.)  Previous studies considered this splitting, which is an effective exchange energy, for six electrons using a frozen-core approximation.\cite{Hu01} We will improve upon these early results both by increasing the number of electrons to include the $N\!=\!10$ and $N\!=\!14$ cases, and by relaxing the assumption that the core electrons are frozen. In order to both test and improve upon the frozen-core approximation, we will use the cutoff approximation as in Sec.~\ref{subsec:convergence}, keeping all multi-particle states that have a non-interacting energy below a certain cutoff excitation energy, $E_c$, from the non-interacting ground state.  When $E_c\!=\!0$, we obviously have only the four multi-particle states which correspond to the four ways of distributing the two valence electrons among the two valence orbitals. As the cutoff is increased, the number of states included quickly increases at a rate which grows with $N$.

\begin{figure}
\subfigure{
\includegraphics[width=.8\columnwidth]{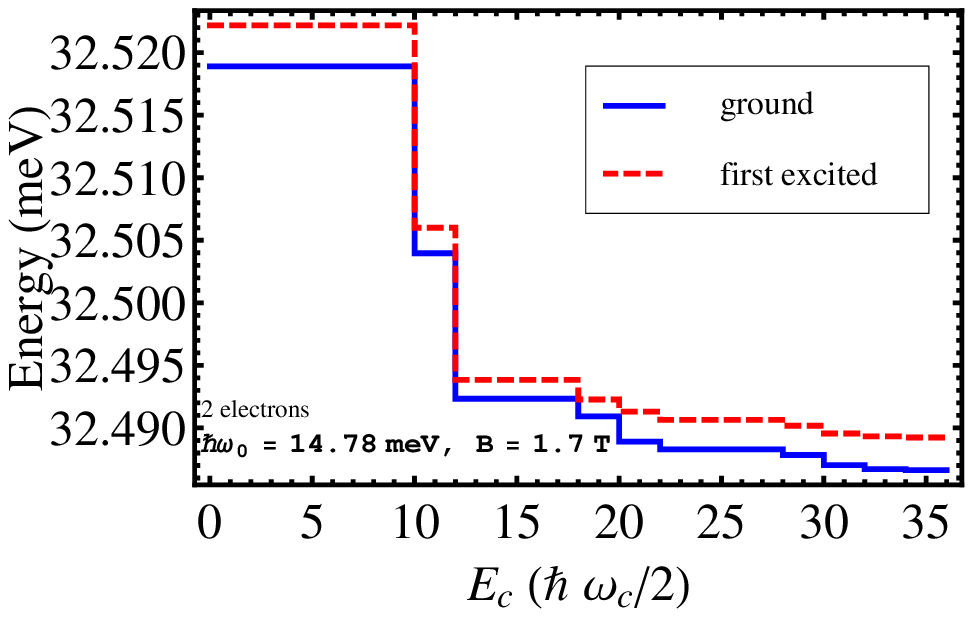}}
\subfigure{
\includegraphics[trim=0mm 0mm 10mm 0mm, clip,width=.8\columnwidth]{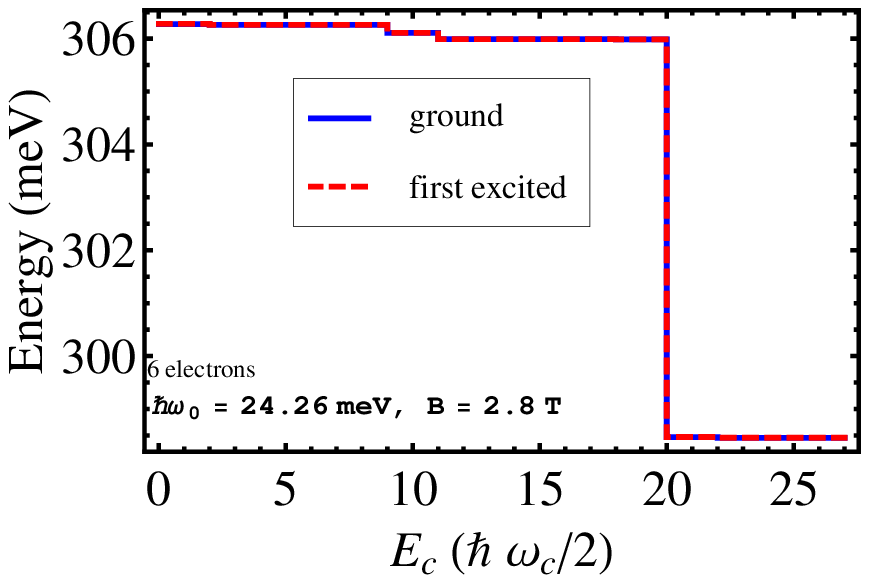}}
\subfigure{
\includegraphics[trim=0mm 0mm 4mm 0mm, clip,width=.8\columnwidth]{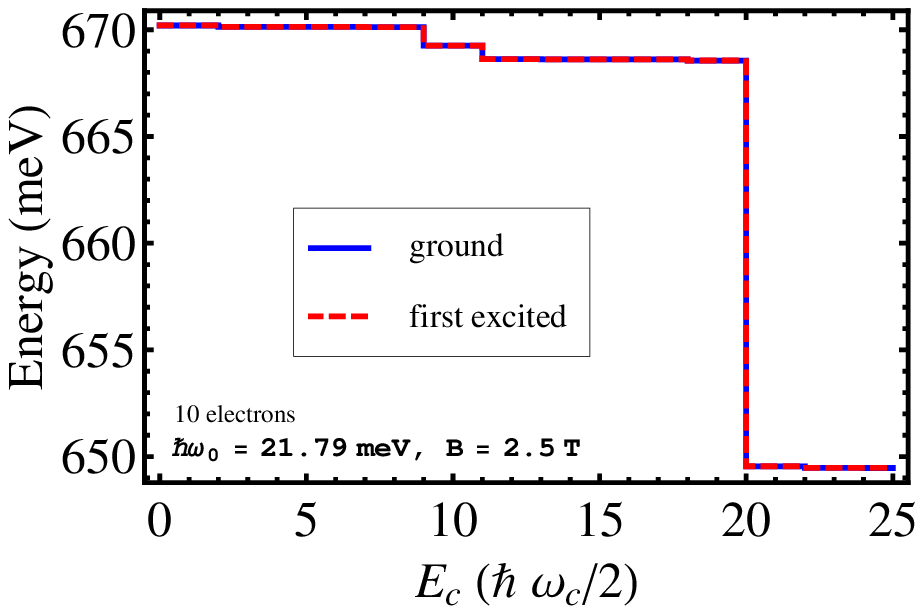}}
\subfigure{
\includegraphics[width=.8\columnwidth]{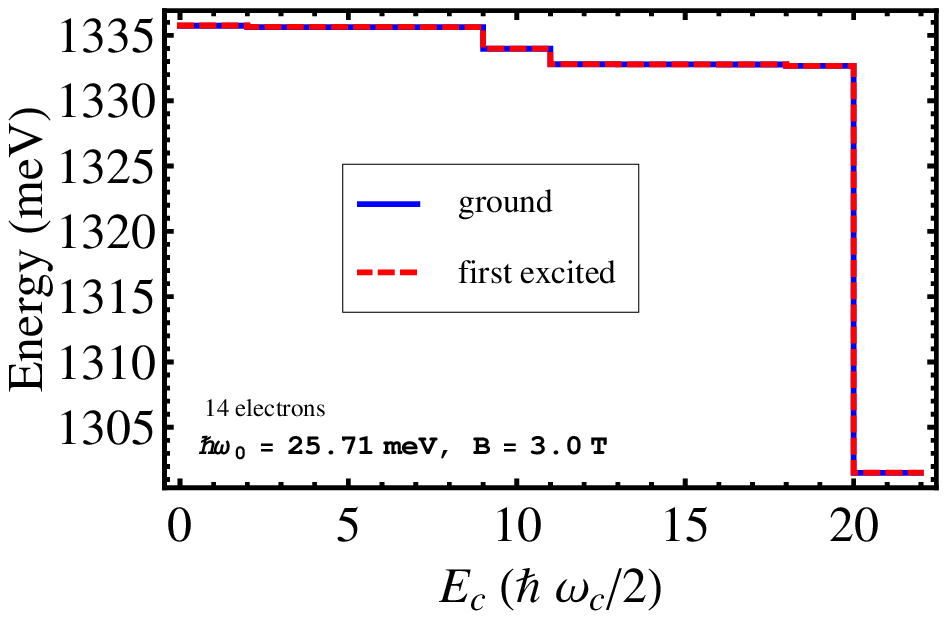}}
\caption{\label{fig:Econvergence} (Color online.) Ground and first excited state energies (left) for 2, 6, 10, and 14 (top to bottom) electrons and the difference between the two (right) vs.~energy cutoff.}
\end{figure}

\begin{figure}
\includegraphics[width=\columnwidth]{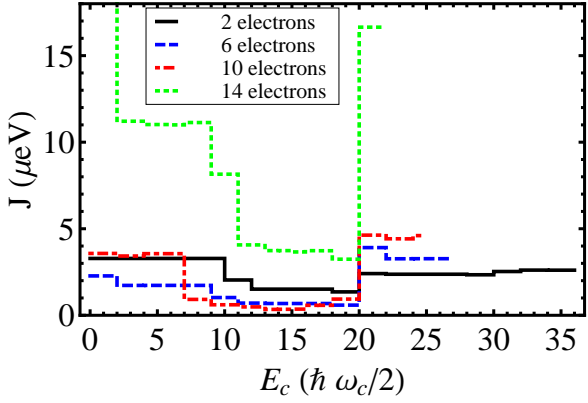}
\caption{(Color online.) Splitting between ground and first excited state energies vs.~energy cutoff for 2, 6, 10, and 14 electrons.}\label{fig:Jconvergence}
\end{figure}

In Fig.~\ref{fig:Econvergence} we show the convergence of the lowest two eigenenergies, and in Fig.~\ref{fig:Jconvergence} we show their difference, $J$. In the two-electron case it appears that $J$ is well converged for cutoffs larger than $E_c \!=\! 20\hbar\omega_c/2$, in agreement with our expectations from the single-dot results of Sec.~\ref{subsec:convergence}. It is also apparent in Figs.~\ref{fig:Econvergence} and \ref{fig:Jconvergence} that a substantial jump occurs at $E_c \!=\! 20\hbar\omega_c/2$ for all numbers of electrons. The significance of this special cutoff value (which depends on our choice of parameters) is that it corresponds to the point at which the lowest excited state with the same magnetic quantum number $m$ as the valence orbital is included in the cutoff scheme. When the valence orbital has quantum numbers $(n,m)$, this newly added state has quantum numbers $(n+2,m)$. Coulomb matrix elements involving this state and the valence orbital are quite large compared with matrix elements between states with different values of $m$.  The reason can be seen by considering a single dot (or equivalently, a DQD with infinite interwell separation).  In that case, angular momentum is a good quantum number.  Thus the ground state must be built from a set of orbitals which all have the same angular momentum.  Thus, if one considers the convergence of ground state energy as orbitals are added to the basis set one-by-one, the energy should only be lowered when an orbital with magnetic quantum number $m$ is added to the basis.  Now, angular momentum is no longer conserved for a coupled DQD, and the ground state contains contributions from all orbitals.  However, for the physically relevant interdot distance we have taken, the ground state still favors the set of orbitals discussed above.  Hence the energy changes more when we add the $(n+2,m)$ orbital than when we add others.

These observations highlight the importance of using cutoff energies above $E_c \!=\! 20\hbar\omega_c/2$. As $E_c$ is increased, it eventually becomes large enough that Fock-Darwin states beyond those kept in the CI calculations become excited, and it would be inconsistent to increase $E_c$ any further since we would be adding higher-energy configurations without first including all configurations with lower energy. For the cases $N\!=\!2,6,10$, it suffices to keep up through the $F$ shell in order to reach the $E_c \!>\! 20\hbar\omega_c/2$ regime, and the upper bound on $E_c$ arising from these considerations is fixed by the energy difference between the valence state and the $G_{--}$ orbital (and so depends on the number of electrons). For the $N\!=\!14$ case, however, it is necessary to include two orbitals from the $G$ shell in order to achieve $E_c \!>\! 20\hbar\omega_c/2$. Thus for $N\!=\!2,6,10,14$ we must have $E_c/\left(\hbar\omega_c/2\right) \!<\! 36,27,25,22$, respectively. Although it is evident from Figs.~\ref{fig:Econvergence} and \ref{fig:Jconvergence} that small fluctuations in the lowest energies and $J$ continue to arise beyond $E_c \!=\! 20\hbar\omega_c/2$, there are indications that no additional large jumps occur as $E_c$ is increased further. In particular, we have noticed that such jumps are also apparent in the single-particle matrix elements. The angular-momentum considerations above would suggest that such jumps would occur at integer multiples of $20\hbar\omega_c/2$; however, we have computed single-particle matrix elements including up through the $n=6$ shell and found that no further jumps arise. Although a similar calculation for multi-particle matrix elements would be computationally very challenging, these single-particle results do at least suggest that the curves in Figs.~\ref{fig:Econvergence} and \ref{fig:Jconvergence} are reasonably well converged.


We have used the parameters from the previous subsection so we can directly compare the exchange energies for different electron numbers at a fixed tunneling rate, $t\!\sim\! 0.2$ meV.  We see from Fig.~\ref{fig:Jconvergence} that, in contrast to the zero magnetic field case within a frozen-core approximation,\cite{Hu01} the exchange energy generally increases for larger numbers of electrons.

\begin{figure}
  \includegraphics[trim=0mm 12mm 0mm 12mm, clip,width=.9\columnwidth]{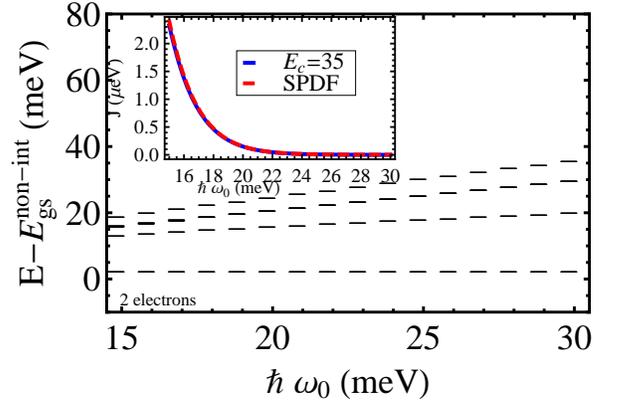}
  \includegraphics[trim=0mm 12mm 0mm 12mm, clip,width=.9\columnwidth]{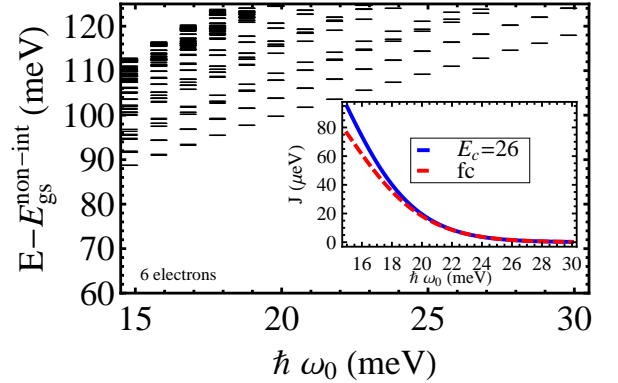}
  \includegraphics[trim=0mm 12mm 0mm 15mm, clip,width=.9\columnwidth]{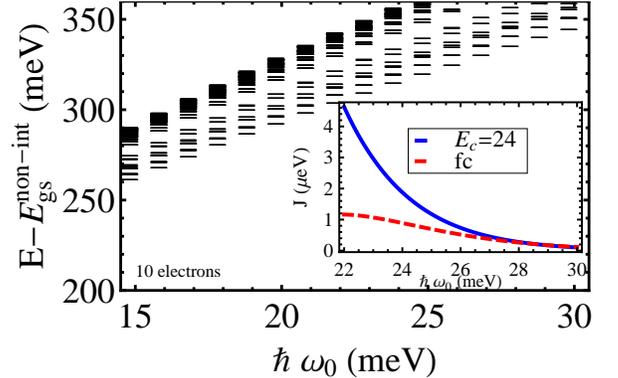}
  \includegraphics[trim=0mm 12mm 0mm 15mm, clip,width=.9\columnwidth]{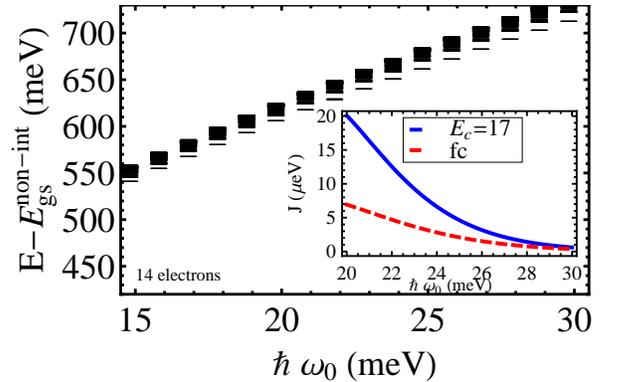}
  \caption{Spectrum vs.~trap frequency for 2, 6, 10, and 14 (top to bottom) electrons. Inset: Exchange energy vs.~trap frequency using cutoff energy and frozen-core approximations for comparison.}\label{fig:Jvsomega}
\end{figure}

Fig.~\ref{fig:Jvsomega} shows how the low-lying spectrum behaves as a function of harmonic well frequency, $\omega_0$, with magnetic field chosen such that $\omega_c/\omega_0 \!=\! 2/\sqrt{99}$ is held constant.  We have subtracted off a trivial linear dependence on $\omega_0$ arising from the non-interacting ground state energy using Eq.~(\ref{eq:Enonint}).  Only the lowest states are shown; in reality there are of course many closely-spaced excited states in the empty upper portion of the plots. Note that, as expected, the lowest pair of states (indistinguishable on the scale shown) are isolated from the other states by a large energy splitting due to the tight confinement and strong interactions.  The singlet-triplet splitting, $J$, is shown in the insets, including results using the frozen-core approximation for comparison.  Generally, the frozen-core approximation is valid in the limit of very large trap frequencies such that the Coulomb interactions become negligible.  However, for realistic dot sizes, the frozen-core restriction underestimates the exchange considerably.  The discrepancy becomes more pronounced as the dot filling is increased.

\subsection{Qubit operation}\label{subsec:dJ}
Finally, in addition to the benefits in fabrication, we now show that the screening is likewise beneficial in the operation of singlet-triplet qubits in the presence of charge noise.  Random telegraph fluctuations in the occupation of a charge impurity center results in fluctuations, $\delta J$, in the exchange energy, inducing singlet-triplet decoherence and gate errors.  Recent experiments have made remarkable progress in countering dephasing due to hyperfine coupling to the nuclear spin bath, extending the decoherence time to $\!\sim\! 100 \mu$s,\cite{Barthel10,Bluhm11} leaving charge fluctuations as the dominant source of error.  This presents a formidable obstacle to current efforts to perform high-fidelity logical gates.  Although one may be able to find ``sweet spots" where $\delta J$ is minimized for a given noise channel,\cite{Stopa08} e.g., fluctuation in interwell detuning, one generally cannot simultaneously protect against fluctuations in the other model parameters.\cite{Nielsen10}  Furthermore, in a large-scale system, the fine-tuning required to ensure that the whole ensemble is operating at a sweet spot would seem to be prohibitive.  On the other hand, using multi-electron DQDs in order to reduce $\delta J$ via screening is a relatively general and straightforward approach.  It is thus desirable to consider the effectiveness of the screening in protecting against noise from charge fluctuators.

\begin{figure}
  \includegraphics[width=\columnwidth]{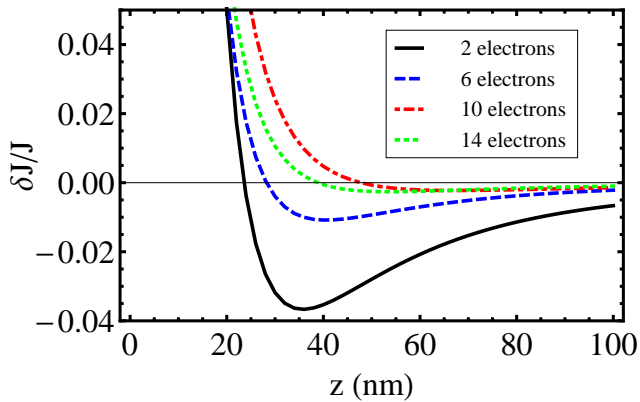}
  \caption{(Color online.) Fractional change in exchange energy due to impurity vs.~impurity distance for different numbers of electrons using the energy cutoff approximation with $E_c\!=\!35,26,24,17$ for $N\!=\!2,6,10,14$ electrons respectively.}\label{fig:dJoverJvsz}
\end{figure}

Decoherence from the charge noise is essentially determined by the average switching time, $\tau$, for all but the most minute values of $\delta J$ on the order of $\hbar/\tau$.  Gate errors, though, are dominated by the relative fluctuation in the exchange energy, $\delta J/J$.  We calculate this quantity for a single repulsive charge impurity center located directly above one of the dots, neglecting any image charge.  This is only meant to give a qualitative picture of gate errors due to charge noise.  Although it would be no more difficult to perform the calculation with any given impurity potential, in the absence of direct knowledge of the impurity positions and switching times relevant to a specific sample, any choice is purely arbitrary. As shown in Fig.~\ref{fig:dJoverJvsz}, the relative fluctuation in $J$ for a multi-electron qubit may be reduced by as much as an order of magnitude compared to the single-electron-per-dot case, which would result in a dramatic reduction of the charge noise induced infidelity. Figure \ref{fig:dJoverJvsz} shows the results of using the best energy cutoff approximation in each case.

\begin{figure*}
  \subfigure[]{\includegraphics[width=.98\columnwidth]{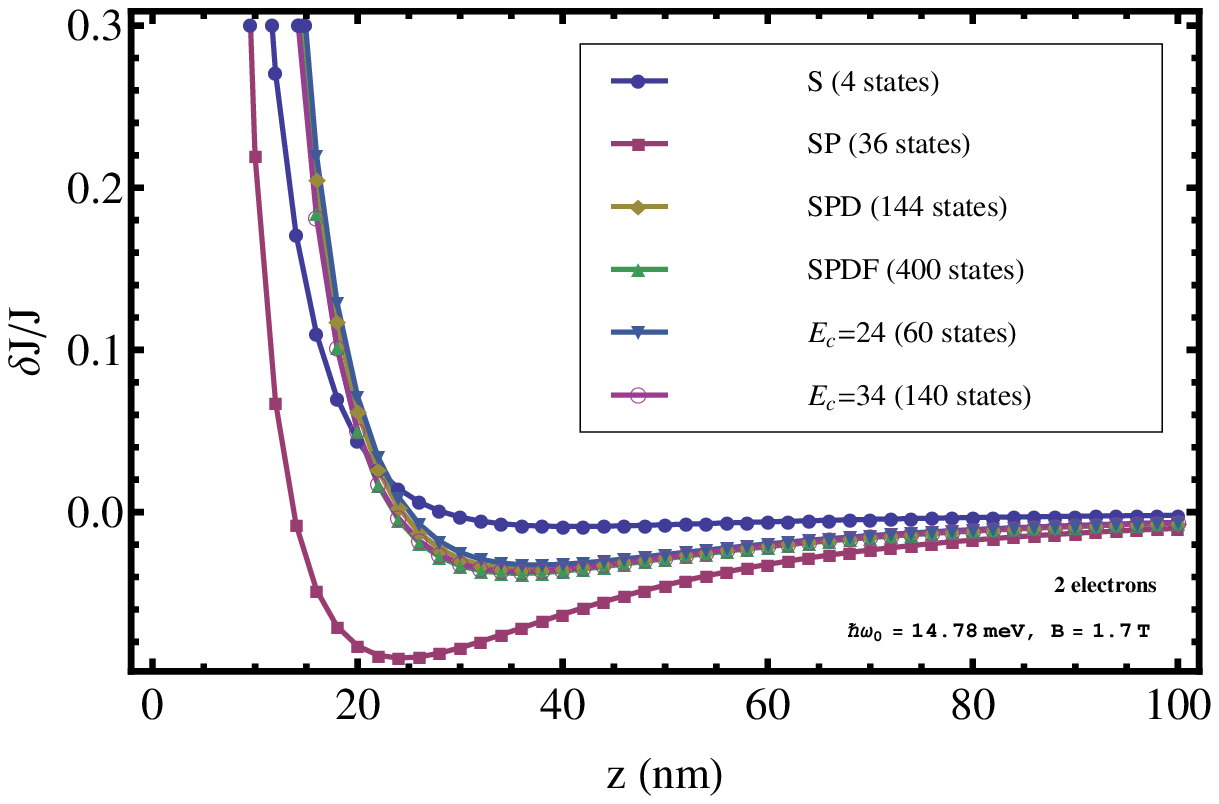}}
  \subfigure[]{\includegraphics[width=.98\columnwidth]{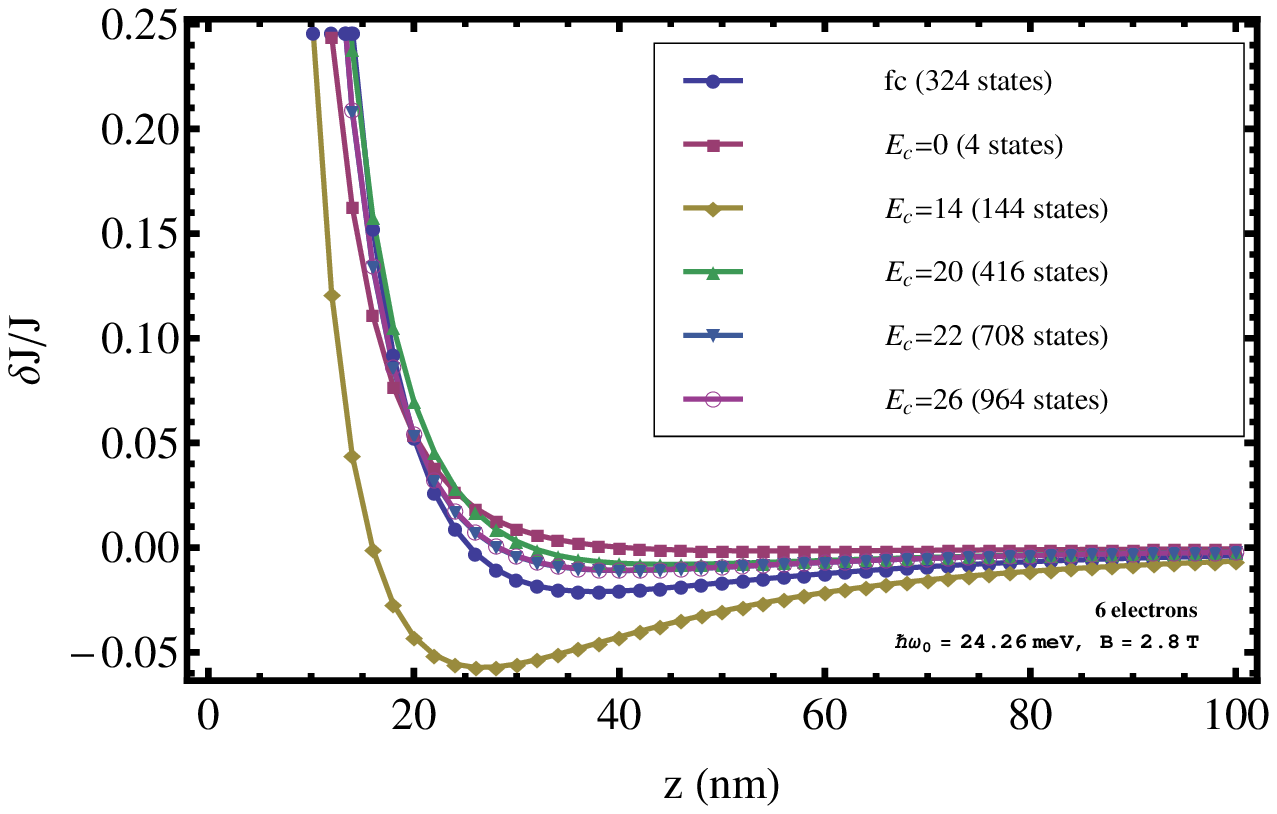}}
  \subfigure[]{\includegraphics[width=.98\columnwidth]{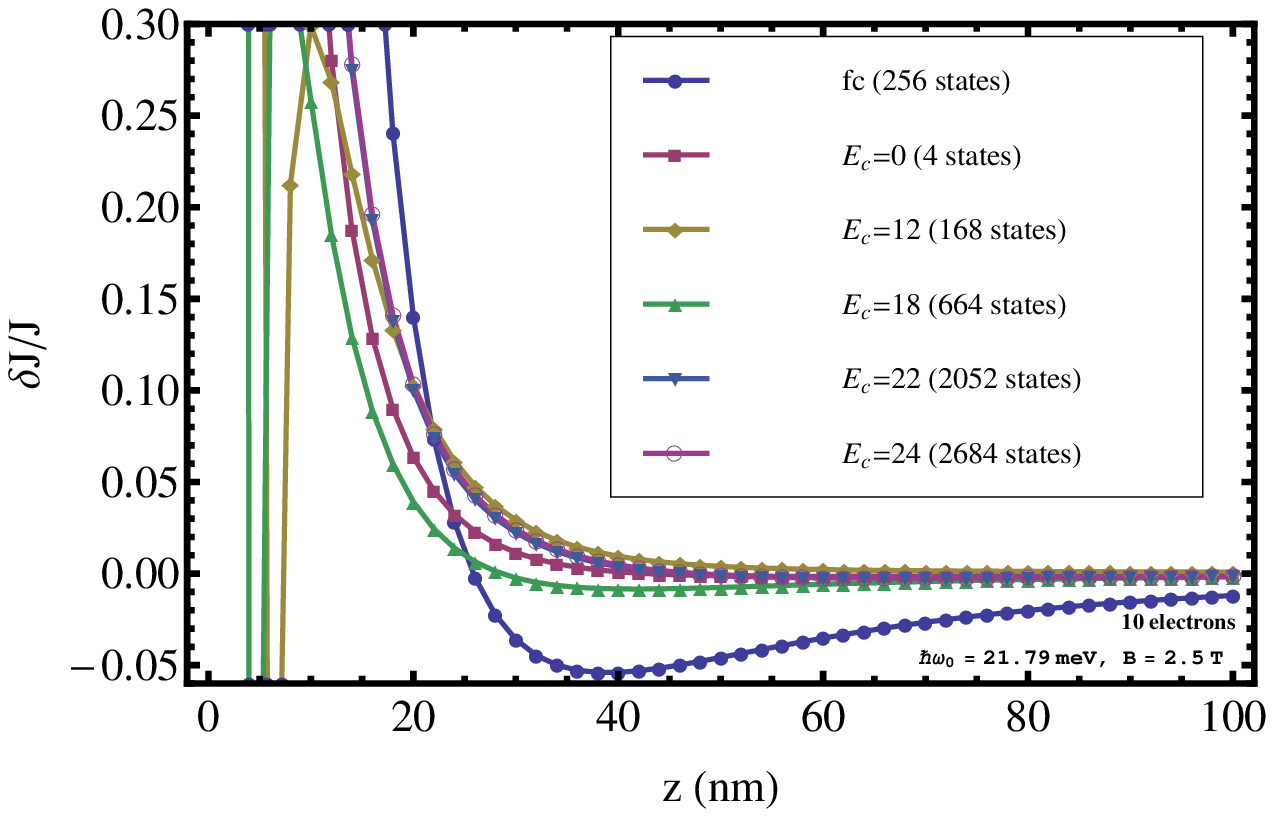}}
  \subfigure[]{\includegraphics[width=.98\columnwidth]{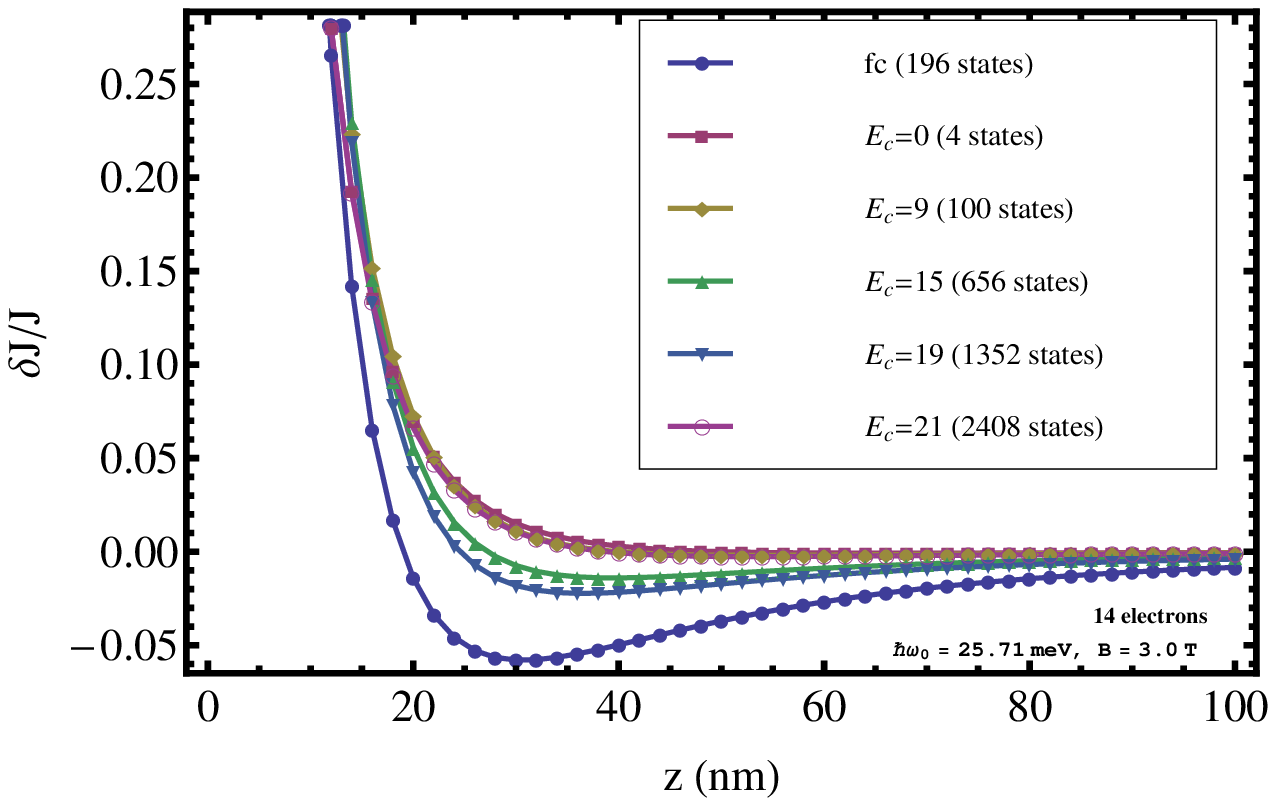}}
  \caption{(Color online.) Fractional change in exchange energy due to impurity vs.~impurity distance for (a) 2, (b) 6, (c) 10, and (d) 14 electrons with various energy cutoffs.}\label{fig:dJoverJvsz_cutoffs}
\end{figure*}

Generally speaking, more electrons correspond to less sensitivity to the charge fluctuator.  In the ten-electron case, though, the sensitivity appears anomalously small.  This may be due to the valence electrons residing in the $P_+$ orbitals, which have a suppressed tunneling rate, as discussed above.  As a result of this orbital effect, the trapping frequency, $\omega_0$, does not have to be increased as much to match the two-electron $S$ orbital tunneling rate.  This in turn implies that the core electrons have more freedom to screen the impurity.

Fig.~\ref{fig:dJoverJvsz_cutoffs} shows the convergence of $\delta J$ as the energy cutoff $E_c$ approaches its maximal value, as well as full CI results for the two-electron case and frozen-core results for the other cases. It is not guaranteed that our results for the exchange energy are fully converged, especially for larger numbers of electrons, but Fig.~\ref{fig:dJoverJvsz_cutoffs} demonstrates that the qualitative trends apparent in Fig.~\ref{fig:dJoverJvsz} are reliable, most notably the decrease in sensitivity to the charged impurity for larger numbers of electrons. These figures also illustrate the worsening of the frozen-core approximation as electrons are added to the system.

\section{Conclusions}\label{sec:conclusions}
The presence of randomly positioned and temporally fluctuating charge impurity centers in even the cleanest semiconductor samples is currently the most prominent roadblock to experimental realization of precisely controlled, long-lived semiconductor spin qubits.  It is an open question to what extent and in what direction the path to scalable spin quantum computation will be altered by this roadblock.  On one hand, one could pursue ultraclean samples with impurity concentration under $10^{12} \text{cm}^{-3}$.\cite{Nguyen11}  On the other hand, one could seek to engineer qubits that are less sensitive to the presence of charge impurities, as we have discussed in this paper.  However, since the standard two-qubit gate for singlet-triplet qubits relies on the Coulomb interaction,\cite{Taylor05,Weperen11} engineering a qubit that is less sensitive to fluctuations in the electrostatic environment also diminishes the ability to manipulate the many-qubit state via the standard techniques.  This is, of course, a familiar conundrum in all proposals of quantum computation.  The outlook is nevertheless quite promising, since random fluctuations in the electrostatic potential can be suppressed by bias cooling\cite{Pioro-Ladriere05} or by adding a negatively biased insulated top gate\cite{Buizert08} in order to suppress leakage of electrons from the gates into the two-dimensional electron gas.  Furthermore, intentional changes in the electrostatic potential for the purposes of two-qubit operations can be enhanced by a floating interdot capacitive gate.\cite{Chan02,Hubel07}  It is therefore reasonable to sacrifice a little of the abundance of controllability in order to gain a measure of immunity to the impurity background.

In this paper, we have explored the ramifications of using multi-electron quantum dots in order to diminish the effect of the random impurity potential.  The numerical calculations in this work provide qualitative guidance to experiments regarding the characteristics of a DQD loaded with $2$, $6$, $10$, or $14$ electrons and how its behavior depends on dot size and nearby charge impurity centers.  Our method of approximation consistently accounts for deviations of the many-body wavefunction from an effective two-electron, frozen-core treatment, and our qualitative observations are shown to be robust and physically sensible.  Pursuing quantitatively more precise results would be fairly meaningless since we are using a simplified model of the confinement potential.  A precise calculation would require a fully self-consistent Schrodinger-Poisson solution for a particular gate geometry and a specific impurity distribution.\cite{Stopa96,Melnikov06}  However, the simpler and more general model potential has previously been found to be a good approximation to the exact one.\cite{Nielsen10}  In any case, the essential screening physics discussed above is a general feature which should not depend sensitively on the exact form of the confinement.  A more important issue is the fact that our calculations were performed in the case of an unusually tight confinement in order to reduce the relative strength of the Coulomb interactions and aid convergence.  This precludes a direct connection to recent experiments using dots several times larger.  However, for larger dots the features we have discussed should become even more pronounced as the deviations from the frozen-core approximation become more important.

In this work, we have characterized multi-electron qubits in the presence of a charged impurity.  We have not discussed details of initialization, manipulation, and readout.  We envision these being performed as in the two-electron case\cite{Petta05} by tilting the double-well so that the valence electrons begin to shift onto the same site, resulting in larger overlap and exchange.  These are separate issues which, although not expected to be problematic, might require some care and could constitute the subject of future investigation. We have demonstrated that using multi-electron DQDs in a finite magnetic field to form singlet-triplet qubits is a viable option that does not suffer from the suppression of the exchange energy that one might expect based on previous results in a more restricted approximation.\cite{Hu01}  Not only are multi-electron qubits viable, they are also an attractive option due to nascent screening of the rough background by the few ``core" electrons in each dot.  This results in reduced sensitivity to random static impurity potentials, allowing easier fabrication, as well as reduced sensitivity to fluctuations in the impurity potential, facilitating more reliable single-qubit manipulation.

This work is supported by LPS-NSA and IARPA.

\appendix

\section{Fock-Darwin states}\label{app:A}
\subsection{One harmonic well}
The Fock-Darwin Hamiltonian has the form
\be
H={1\over2m^*}\left(-i\hbar \mathbf{\partial}+{e\over c} \mathbf{A}\right)^2+{1\over2}m^*\omega_0^2 r^2,\label{hami}
\ee
where $r=\sqrt{x^2+y^2}$ is the 2d radial coordinate, $m^*$ is the effective electron mass, and the vector potential is given by
\be
\mathbf{A}=-{B\over2}y\mathbf{\hat x}+{B\over2}x\mathbf{\hat y}.\label{vecpot}
\ee
Introducing the cyclotron frequency,
\be
\omega_c\equiv {eB\over m^*c},
\ee
we can write the Hamiltonian as
\be
H=-{\hbar^2\over2m^*}\nabla^2+{1\over2}m^*\left(\omega_0^2+\omega_c^2/4\right)r^2+{\omega_c\over2}\ell_z,
\ee
where $\ell_z$ is the $z$-component of the angular momentum:
\be
\ell_z=-i\hbar(x\partial_y-y\partial_x).
\ee
It will be useful to consider the Fock-Darwin Hamiltonian in dimensionless complex coordinates, defined by:
\begin{multline}
z = {x+iy\over\sqrt{2}\ell_0}, \quad \bar{z} = {x- i y\over\sqrt{2}\ell_0},
\\
\partial = {\ell_0\over\sqrt{2}}\left(\partial_x-i\partial_y\right), \quad \bar\partial = {\ell_0\over\sqrt{2}}\left(\partial_x+i\partial_y\right),
\end{multline}
where the length scale $\ell_0$ is
\be
\ell_0\equiv \ell_B\left(1/4+\omega_0^2/\omega_c^2\right)^{-1/4},
\ee
and the magnetic length is given by
\be
\ell_B=\sqrt{\hbar c\over eB}.
\ee
Defining the following set of creation and annihilation operators,
\begin{multline}
a={1\over\sqrt{2}}(\bar{z}+\partial),\quad
a^\dagger={1\over\sqrt{2}}(z-\bar\partial),
\\
b={1\over\sqrt{2}}(z+\bar\partial),\quad
b^\dagger={1\over\sqrt{2}}(\bar{z}-\partial),
\end{multline}
which have the commutation relations
\be
[a,a^\dagger]=1,\quad [b,b^\dagger]=1,\label{comrels}
\ee
with all other commutators vanishing, the Hamiltonian becomes
\be
H=\hbar\omega_+(a^\dagger a+1/2)+\hbar\omega_-(b^\dagger b+1/2),
\ee
with
\be
\omega_\pm\equiv \sqrt{\omega_0^2+\omega_c^2/4}\pm\omega_c/2.
\ee
The ground state of this Hamiltonian lives in the kernel of both $a$ and $b$ and so has the (normalized) wavefunction
\be
\psi_0 = \sqrt{2\over\pi}e^{-z\bar z}.
\ee
It is trivial to check that $a\psi_0=b\psi_0=0$. The commutation relations (\ref{comrels}) imply that
\bea
a^\dagger a(a^\dagger)^n &=& n(a^\dagger)^n + (a^\dagger)^{n+1}a,\nn\\
b^\dagger b(b^\dagger)^n &=& n(b^\dagger)^n + (b^\dagger)^{n+1}b,
\eea
so that the class of (normalized) functions
\be
\psi_{n_+,n_-}={1\over\sqrt{n_+!n_-!}}(a^\dagger)^{n_+}(b^\dagger)^{n_-}\psi_0
\ee
are eigenfunctions of the Hamiltonian with eigenvalues
\be
E_{n_+,n_-}=\hbar\omega_+(n_++1/2)+\hbar\omega_-(n_-+1/2).
\ee
The quantum numbers $n_+$ and $n_-$ are nonnegative integers. If we define
\be
n\equiv n_++n_-,\qquad m\equiv n_+-n_-,
\ee
then $n$ is a nonnegative integer, and $m$ takes values from $-n$ to $n$ in steps of 2. The spectrum in terms of $n$ (principal quantum number) and $m$ (magnetic quantum number) is
\be
E_{n,m}=(n+1)\hbar\sqrt{\omega_0^2+\omega_c^2/4}+m\hbar\omega_c/2.\label{spectrum}
\ee

\subsection{Two wells located at $(x,y)=(\pm x_0,0)$}
In the case of two quantum dots, we need to modify slightly the Fock-Darwin states found above. Obviously, the above results are valid no matter where the dot is located so long as our coordinates are defined with respect to the center of the dot. In the case of two lateral quantum dots, however, we want to define the coordinates with respect to the point halfway in between the dots. In these coordinates, the dots are located at $(x,y)\!=\!(\pm x_0,0)$. One might be tempted to implement this coordinate shift simply by defining new complex coordinates:
\be
z_\pm\equiv x\pm x_0+iy,\qquad \bar z_\pm\equiv x\pm x_0-iy.\label{newz}
\ee
However, this redefinition moves not only the location of the quantum dot but also changes the vector potential $\mathbf{A}$, Eq. (\ref{vecpot}). We can correct this by also performing a gauge transformation on $\mathbf{A}$:
\be
\mathbf{A}\to \mathbf{A}\mp{B\over 2}x_0\mathbf{\hat y},
\ee
which enables us to keep $\mathbf{A}$ fixed to the form in Eq. (\ref{vecpot}) while we move the harmonic well in the $x$-direction. Of course, a gauge transformation affects not only the vector potential but also the wavefunction of the electron, and in particular does so in such a way that the wavefunction picks up a phase:
\be
\psi\to e^{i\varphi}\psi.
\ee
The phase $\varphi$ can be fixed by going back to Eq. (\ref{hami}), implementing the gauge transformation on $\mathbf{A}$, and demanding that the new wavefunction is still an eigenfunction of the Hamiltonian, leading to the condition
\be
\left(-i\hbar\partial\mp{e\over c}{B\over2}x_0 \mathbf{\hat y}\right)e^{i\varphi}=0.
\ee
This condition has the solution
\be
\varphi_\pm=\pm {e B x_0\over 2\hbar c}y,\label{phase}
\ee
where the $\pm$ corresponds to the dot at $x\!=\!\mp x_0$, or in other words, the right dot has the minus sign in the phase, the left dot has the plus. Notice that this phase is the same for all states living in the same dot. The bottom line is that we may continue to use the same Fock-Darwin states as before, but with $z$ and $\bar z$ now defined according to Eq. (\ref{newz}) and with the overall phase factor $e^{i\varphi_\pm}$ included, where $\varphi_\pm$ is given in Eq. (\ref{phase}). Also note that, when using the raising operators to generate states, the phase is to be included {\sl after} the operators have been applied to the phaseless $\psi_0$:
\be
\psi^\pm_{n_+,n_-}={\sqrt{2}\over\sqrt{\pi n_+!n_-!}}e^{i\varphi_\pm}(a_\pm^\dagger)^{n_+}(b_\pm^\dagger)^{n_-}e^{-z_\pm\bar z_\pm}.\label{fdstates}
\ee
Here, the $+(-)$ index gives a state in the left (right) dot. We have taken the liberty of defining the operators $a^\dagger_\pm$ and $b^\dagger_\pm$, which are just the usual creation operators but with $z$ and $\bar z$ replaced by $z_\pm$ and $\bar z_\pm$. In terms of $n$ and $m$ quantum numbers, we have
\be
\phi_{nm}^\pm(x,y)\equiv\psi^\pm_{{n+m\over2},{n-m\over2}}(x,y).
\ee

\section{One and two-particle matrix elements}\label{app:B}
\subsection{Integrals}
Matrix elements involving products of Fock-Darwin states can be simplified by exploiting the fact that all such states are generated from $e^{-z\bar z}$. (In this section, we suppress the $\pm$ indices indicating to which dot the state belongs.) Since $a^\dagger$ and $b^\dagger$ are linear differential operators acting on $e^{-z\bar z}$, the Fock-Darwin wavefunctions have the form $P(z,\bar z)e^{-z\bar z}$ where $P(z,\bar z)$ is some bivariate polynomial. This in turn means that we can also generate the states by instead starting from the generator
\be
\phi(z,\bar z,c,d)=\sqrt{2\over\pi}e^{-z\bar z+cz+d\bar z}
\ee
and differentiating with respect to $c$ and $d$. The explicit form of the differential operator which yields the Fock-Darwin states is
\begin{widetext}
\be
D^{n_+,n_-}_{c,d}=
\left\{
  \begin{array}{ll}
    \sqrt{\frac{n_+!}{n_-!}}(-1)^{n_-}2^{(n_+-n_-)/2}\partial_c^{n_+-n_-}\sum_{k=0}^{n_-}
\binom{n_-}{k}\frac{(-2)^k}{(n_+-n_-+k)!}\partial_c^k \partial_d^k, & \hbox{$n_+\ge n_-$} \\
    \sqrt{\frac{n_-!}{n_+!}}(-1)^{n_+}2^{(n_--n_+)/2}\partial_d^{n_--n_+}\sum_{k=0}^{n_-}\binom{n_+}{ k}\frac{(-2)^k}{(n_--n_++k)!}\partial_d^k\partial_c^k, & \hbox{$n_+< n_-$}
  \end{array}
\right.
\ee
\end{widetext}
and in terms of this operator, the states are
\be
\psi_{n_+,n_-}(z,\bar z)=D^{n_+,n_-}_{c,d}\phi(z,\bar z,c,d)\Bigg|_{\substack{c=\alpha\\d=-\alpha}},
\ee
with
\be
\alpha={\varphi_\pm\over2y}=\pm {e B x_0\over 4\hbar c}.
\ee
(On the right-hand side of this expression, $c$ is speed of light and not the parameter appearing in $\phi(z,\bar z)$.) Now consider an arbitrary single-particle matrix element comprised of some operator $\cal A$ sandwiched between two Fock-Darwin states:
\begin{widetext}
\be
\langle n_+,n_-| {\cal A} |n_+',n_-'\rangle = \int\!\!\!\!\int dx dy \psi_{n_+,n_-}^*(z,\bar z){\cal A}\psi_{n_+',n_-'}(z,\bar z) = D^{n_+,n_-}_{c,d}D^{n_+',n_-'}_{c',d'}\int\!\!\!\!\int dx dy \phi(\bar z,z,c,d){\cal A}\phi(z,\bar z,c',d')\Bigg|_{\substack{c=\alpha\\d=-\alpha\\ c'=\alpha'\\d'=-\alpha'}}.\label{spME}
\ee
\end{widetext}
We see that we can compute all such matrix elements for an operator $\cal A$ by first computing the object
\be
\int \int dx dy \phi(\bar z,z,c,d){\cal A}\phi(z,\bar z,c',d'),
\ee
applying the differential operators $D^{n_+,n_-}_{c,d}$ and $D^{n_+',n_-'}_{c',d'}$, and then finally setting $c,d,c',d'$ appropriately. Note that we have swapped the $z$ and $\bar z$ in $\phi(\bar z,z,c,d)$ because this generator corresponds to the complex-conjugated wavefunction. Since the parameters $c$ and $d$ are always real numbers, complex conjugation is implemented simply by swapping $z$ and $\bar z$ in the generator $\phi(z,\bar z)$. The utility of this approach is that we need only compute the double integral once and for the simplest possible wavefunctions---pure Gaussians. We have outlined the approach for single-particle matrix elements, but the same procedure also applies to two-particle matrix elements such as we have for the Coulomb interactions between electrons.

Each term of the double well Hamiltonian can be computed in the manner outlined above. Fortunately, the integral for the double harmonic potential can be computed exactly. The integrals over $x$ and $y$ can also be computed for the impurity terms once the $1/r$ Coulomb potential is rewritten as an exponential using the following identity:
\be
{1\over r}={1\over\sqrt{\pi}}\int_0^\infty {ds\over\sqrt{s}}e^{-sr^2}.
\ee
This relation also allows the four coordinate integrations to be performed in the case of the Coulomb-interaction terms. However, for both the Coulomb terms and the impurity terms, the remaining integration over $s$ from the above identity must be done numerically for each set of quantum numbers since we do not have a closed form expression for this integral as a function of the parameters $c$ and $d$.

There are several ways that one could treat the kinetic terms. In order to compute these terms, we need to first act with the kinetic operator on $\phi(z,\bar z,c,d)$ as shown schematically in Eq. (\ref{spME}). Since the kinetic operator acts only on $\phi(z,\bar z,c,d)$, we can rewrite it as a differential operator acting on $c$ and $d$ instead of on $z$ and $\bar z$. Explicitly, it has the form
\begin{multline}
{\cal K}_\pm = -{2\hbar\omega_+\omega_-\over\omega_++\omega_-}\partial_d\partial_c+\hbar\omega_+\left[c\mp {x_0\over\sqrt{2}\ell_0}{\omega_+-\omega_-\over\omega_++\omega_-}\right]\partial_c
\\
+\hbar\omega_-\left[d\pm {x_0\over\sqrt{2}\ell_0}{\omega_+-\omega_-\over\omega_++\omega_-}\right]\partial_d+{\hbar\over2}(\omega_++\omega_-)(1-cd)
\\
+{x_0^2\hbar\over4\ell_0^2}{(\omega_+-\omega_-)^2\over\omega_++\omega_-} \mp{x_0\hbar\over2\sqrt{2}\ell_0}(\omega_+-\omega_-)(c-d).
\end{multline}

\subsection{Symmetries}

In computing single and two-particle matrix elements, there are a few symmetries that we may exploit to reduce the number of matrix elements that need to be computed. First of all, each single-particle term in the Hamiltonian is real and symmetric:
\be
\langle \eta,n,m|{\cal A}| \eta',n',m'\rangle = \langle \eta',n',m'|{\cal A}| \eta,n,m\rangle.
\ee
We have included the parameter $\eta\!=\!\pm1$ in the states to designate whether the state is associated with the right dot ($\eta\!=\!-1$) or the left ($\eta\!=\!+1$). As far as the impurity terms go, this is the only symmetry they obey, at least for an impurity located away from $x\!=\!0,y\!=\!0$. The kinetic and double well potential terms, however, also preserve a left-right symmetry which mixes with the single-dot parity operator $(-1)^{n+n'}$:
\be
\langle \eta,n,m|{\cal A}| \eta',n',m'\rangle = (-1)^{n+n'}\langle -\eta,n,m|{\cal A}|-\eta',n',m'\rangle.
\ee
The two-particle Coulomb matrix elements form a subgroup of the permutation group:
\begin{widetext}
\begin{multline}
\langle \eta_1,n_1,m_1;\eta_2,n_2,m_2|{\cal C}|\eta_1',n_1',m_1';\eta_2',n_2',m_2'\rangle= \langle \eta_2,n_2,m_2;\eta_1,n_1,m_1|{\cal C}|\eta_2',n_2',m_2';\eta_1',n_1',m_1'\rangle \\=\langle \eta_1',n_1',m_1';\eta_2',n_2',m_2'|{\cal C}|\eta_1,n_1,m_1;\eta_2,n_2,m_2\rangle = \langle \eta_2',n_2',m_2';\eta_1',n_1',m_1'|{\cal C}|\eta_2,n_2,m_2;\eta_1,n_1,m_1\rangle.
\end{multline}

The Coulomb matrix elements also exhibit left-right symmetry laced with parity:
\be
\langle \eta_1,n_1,m_1;\eta_2,n_2,m_2|{\cal C}|\eta_1',n_1',m_1';\eta_2',n_2',m_2'\rangle = (-1)^{n_1+n_2+n_1'+n_2'}\langle -\eta_1,n_1,m_1;-\eta_2,n_2,m_2|{\cal C}|-\eta_1',n_1',m_1';-\eta_2',n_2',m_2'\rangle.
\ee
\end{widetext}

\subsection{Orthonormal basis}

Once we have the single and two-particle matrix elements for the various terms in the Hamiltonian, it is useful to switch to an orthonormal basis so that we may subsequently apply the standard Slater-Condon rules (reviewed in Appendix \ref{app:C}) to obtain multi-electron matrix elements. There must exist a linear transformation which takes an operator $\cal A$ in the original basis to ${\cal A}'$ in the orthonormal basis:
\be
{\cal A}' = L{\cal A}L^\dagger,
\ee
where $L^\dagger$ appears to ensure the hermiticity of ${\cal A}'$. Consider now the overlap operator $\cal O$:
\be
\langle\eta,n,m|{\cal O}|\eta',n',m'\rangle=\langle\eta,n,m|\eta',n',m'\rangle.
\ee
If we apply the transformation to $\cal O$, then we must obtain the identity operator by definition:
\be
{\cal O}'=L{\cal O}L^\dagger=1,
\ee
implying that
\be
{\cal O}=L^{-1}(L^{-1})^\dagger.
\ee
This factorization of $\cal O$ is called the Cholesky Decomposition. The components of $L^{-1}$ in the original (non-orthonormal) basis form a lower triangular matrix, and they can be found in a systematic way. The fact that $L^{-1}$ is triangular also fits with the fact that it is the transformation which takes non-orthonormal states to orthonormal states. To see this, consider a matrix element of an operator $\cal A$ between two non-orthonormal states:
\begin{multline}
\langle NO_1|{\cal A}|NO_2\rangle=\langle NO_1|L^{-1}L{\cal A}L^\dagger (L^\dagger)^{-1}|NO_2\rangle
\\
=\langle NO_1|L^{-1}{\cal A}' (L^\dagger)^{-1}|NO_2\rangle.
\end{multline}
From this, it is clear that the orthonormal states are obtained by applying $(L^\dagger)^{-1}$ to non-orthonormal states:
\be
|O\rangle = (L^\dagger)^{-1}|NO\rangle.
\ee

In summary, to switch to the orthonormal basis, we first compute the matrix of overlaps in the non-orthonormal basis and then perform a Cholesky Decomposition on the result to obtain $L$. For each single-particle term ${\cal A}$ in the Hamiltonian, we compute $L{\cal A}L^\dagger$ to obtain $\cal A'$ in the orthonormal basis. For the two-particle (Coulomb) terms, we use a slight generalization of the transformation to the case of multi-index tensors:
\be
{\cal C}'_{\alpha\beta\gamma\delta}=\sum_{\alpha'\beta'\gamma'\delta'}L_{\alpha\alpha'}L_{\beta\beta'}{\cal C}_{\alpha'\beta'\gamma'\delta'}L^\dagger_{\gamma'\gamma}L^\dagger_{\delta'\delta}.
\ee
Here, the index $\alpha$ represents a distinct set of single-particle state quantum numbers, $\alpha\!=\!\{\eta,n,m\}$, and similarly for $\beta$, etc.

\section{Slater-Condon rules for multi-particle matrix elements}\label{app:C}

Once we have all the matrix elements for each term in the Hamiltonian in the orthonormal basis, we can apply standard Slater-Condon rules to obtain matrix elements of the various terms with respect to fully antisymmetrized multi-electron wavefunctions. Since we are interested in multi-particle states that have net spin zero, we can write a generic multi-electron state as follows:
\be
|\Psi\rangle=|\{\alpha^{(1)}_\uparrow,\alpha^{(2)}_\uparrow,...\},\{\alpha^{(1)}_\downarrow,\alpha^{(2)}_\downarrow,...\}\rangle,
\ee
where for example, the symbol $\alpha^{(k)}_\uparrow$ represents a spin-up electron in the Fock-Darwin state with quantum numbers $\alpha^{(k)}\!=\!\{\eta_k,n_k,m_k\}$. We define $|\Psi\rangle$ to be a fully antisymmetrized state, so the $\alpha^{(k)}_\uparrow$, $\forall k$, must be distinct from each other, and likewise for the $\alpha^{(k)}_\downarrow$. Furthermore, the particular order of the $\alpha^{(k)}_\uparrow$ (and similarly for the $\alpha^{(k)}_\downarrow$) does not matter. Therefore, we may pick a canonical ordering, and for concreteness we choose this ordering to be determined by the energy of the single-particle states, with lowest energy states to the left (smaller $k$) (see Eq. (\ref{spectrum})). Obviously, states which only differ by the choice of $\eta$ (which dot they belong to) will have the same energy, so we furthermore specify that in these cases, the state with $\eta\!=\!-1$ resides to the left of the state with $\eta\!=\!1$.

We first consider the single-particle terms in the Hamiltonian, which include the kinetic, double well potential, and impurity terms. Denoting these terms collectively by the operator $\cal A$, we consider the following matrix element,
\be
\langle \Psi'|{\cal A}|\Psi\rangle.
\ee
If $|\Psi\rangle=|\Psi'\rangle$, then according to the Slater-Condon rules, we have
\begin{widetext}
\be
\langle \Psi|{\cal A}|\Psi\rangle = \langle\{\alpha^{(1)}_\uparrow,\alpha^{(2)}_\uparrow,...\},\{\alpha^{(1)}_\downarrow,\alpha^{(2)}_\downarrow,...\}|{\cal A}|\{\alpha^{(1)}_\uparrow,\alpha^{(2)}_\uparrow,...\},\{\alpha^{(1)}_\downarrow,\alpha^{(2)}_\downarrow,...\}\rangle= \sum_{\kappa\in\{\alpha^{(1)}_\uparrow,...\}} \langle \kappa|{\cal A}|\kappa\rangle + \sum_{\kappa\in\{\alpha^{(1)}_\downarrow,...\}} \langle \kappa|{\cal A}|\kappa\rangle.
\ee
\end{widetext}
If $|\Psi\rangle$ and $|\Psi'\rangle$ differ by only one single-particle state, either in the set $\{\alpha^{(1)}_\uparrow,\alpha^{(2)}_\uparrow,...\}$ or in $\{\alpha^{(1)}_\downarrow,\alpha^{(1)}_\downarrow,...\}$, then for example (assuming the states that differ have spin up)
\begin{widetext}
\be
\langle \Psi'|{\cal A}|\Psi\rangle = \langle\{\alpha'^{(1)}_\uparrow,...,\alpha'^{(\ell')}_\uparrow,...\},\{\alpha^{(1)}_\downarrow,...\}|{\cal A}|\{\alpha^{(1)}_\uparrow,...,\alpha^{(\ell)}_\uparrow,...\},\{\alpha^{(1)}_\downarrow,...\}\rangle=(-1)^{\ell+\ell'}\langle\alpha'^{(\ell')}_\uparrow|{\cal A}|\alpha^{(\ell)}_\uparrow\rangle,
\ee
\end{widetext}
where $\alpha^{(\ell)}_\uparrow$ is {\sl not} in the set $\{\alpha'^{(1)}_\uparrow,...,\alpha'^{(\ell')}_\uparrow,...\}$ and $\alpha'^{(\ell')}_\uparrow$ is {\sl not} in the set $\{\alpha^{(1)}_\uparrow,...,\alpha^{(\ell)}_\uparrow,...\}$.
If $|\Psi\rangle$ and $|\Psi'\rangle$ differ by more than one single-particle state, then the matrix element vanishes.

The story is a little more complicated in the case of two-particle operators, which for us only includes the Coulomb interactions. For the diagonal matrix elements, we have
\begin{widetext}
\be
\langle \Psi|{\cal C}|\Psi\rangle = \langle \{\alpha^{(1)}_\uparrow,...\},\{\alpha^{(1)}_\downarrow,...\}|{\cal C}|\{\alpha^{(1)}_\uparrow,...\},\{\alpha^{(1)}_\downarrow,...\}\rangle = \!\!\!\!\!\!\!\! \sum_{\kappa,\lambda\in\{\alpha^{(1)}_\uparrow,...\}\cup\{\alpha^{(1)}_\downarrow,...\}}\!\!\!\!\!\!\!\!\!\!\!\!\!\!\!\!\!\!
\langle \kappa,\lambda|{\cal C}|\kappa,\lambda\rangle - \!\!\!\!\!\!\!\! \sum_{\kappa,\lambda\in\{\alpha^{(1)}_\uparrow,...\}} \!\!\!\!\!\!\!\! \langle \kappa,\lambda|{\cal C}|\lambda,\kappa\rangle - \!\!\!\!\!\!\!\! \sum_{\kappa,\lambda\in\{\alpha^{(1)}_\downarrow,...\}} \!\!\!\!\!\!\!\! \langle \kappa,\lambda|{\cal C}|\lambda,\kappa\rangle.
\ee
If $|\Psi\rangle$ and $|\Psi'\rangle$ differ by one single-particle state (and assuming this state has spin up), we have
\bea
\langle \Psi'|{\cal C}|\Psi\rangle &=& \langle\{\alpha'^{(1)}_\uparrow,...,\alpha'^{(\ell')}_\uparrow,...\},\{\alpha^{(1)}_\downarrow,...\}|{\cal C}|\{\alpha^{(1)}_\uparrow,...,\alpha^{(\ell)}_\uparrow,...\},\{\alpha^{(1)}_\downarrow,...\}\rangle \nn\\
&=& (-1)^{\ell+\ell'} \!\!\!\!\!\!\!\! \sum_{\kappa\in(\{\alpha'^{(1)}_\uparrow,...\}\cap\{\alpha^{(1)}_\uparrow,...\})\cup\{\alpha^{(1)}_\downarrow,...\}} \!\!\!\!\!\!\!\!  \langle \kappa,\alpha'^{(\ell')}_\uparrow|{\cal C}|\kappa,\alpha^{(\ell)}_\uparrow\rangle
-(-1)^{\ell+\ell'} \!\!\!\!\!\!\!\!  \sum_{\kappa\in\{\alpha'^{(1)}_\uparrow,...\}\cap\{\alpha^{(1)}_\uparrow,...\}} \!\!\!\!\!\!\!\!  \langle \kappa,\alpha'^{(\ell')}_\uparrow|{\cal C}|\alpha^{(\ell)}_\uparrow,\kappa\rangle.
\eea
\end{widetext}
In the case of two-particle operators like the Coulomb interaction, if $|\Psi\rangle$ and $|\Psi'\rangle$ differ by two single-particle states, then the result does not vanish. Furthermore, the result will depend on whether the two single-particle states have the same spin or not. We first consider the case where they do have the same spin, and suppose for concreteness that they both have spin up. Then
\begin{widetext}
\bea
\langle \Psi'|{\cal C}|\Psi\rangle &=& \langle\{\alpha'^{(1)}_\uparrow,...,\alpha'^{(\ell_1')}_\uparrow,...,\alpha'^{(\ell_2')}_\uparrow,...\},\{\alpha^{(1)}_\downarrow,...\}|{\cal C}|\{\alpha^{(1)}_\uparrow,...,\alpha^{(\ell_1)}_\uparrow,...,\alpha^{(\ell_2)}_\uparrow,...\},\{\alpha^{(1)}_\downarrow,...\}\rangle \nn \\&=&
(-1)^{\ell_1+\ell_2+\ell_1'+\ell_2'}\left[\langle \alpha^{(\ell_1')}_\uparrow,\alpha^{(\ell_2')}_\uparrow|{\cal C}|\alpha^{(\ell_1)}_\uparrow,\alpha^{(\ell_2)}_\uparrow\rangle - \langle \alpha^{(\ell_1')}_\uparrow,\alpha^{(\ell_2')}_\uparrow|{\cal C}|\alpha^{(\ell_2)}_\uparrow,\alpha^{(\ell_1)}_\uparrow\rangle\right].
\eea
\end{widetext}
A similar relation holds for the case where both pairs of distinct single-particle states have spin down. When the pairs have opposite spin, we instead have
\begin{widetext}
\bea
\langle \Psi'|{\cal C}|\Psi\rangle &=& \langle\{\alpha'^{(1)}_\uparrow,...,\alpha'^{(\ell_1')}_\uparrow,...\},\{\alpha'^{(1)}_\downarrow,...,\alpha'^{(\ell_2')}_\downarrow,...\}|{\cal C}|\{\alpha^{(1)}_\uparrow,...,\alpha^{(\ell_1)}_\uparrow,...\},\{\alpha^{(1)}_\downarrow,...,\alpha^{(\ell_2)}_\downarrow,...\}\rangle \nn\\&=&
(-1)^{\ell_1+\ell_2+\ell_1'+\ell_2'}\langle \alpha'^{(\ell_1')}_\uparrow,\alpha'^{(\ell_2')}_\downarrow|{\cal C}|\alpha^{(\ell_1)}_\uparrow,\alpha^{(\ell_2)}_\downarrow\rangle.
\eea
\end{widetext}


\begin{thebibliography}{99}
\bibitem{note} At high numbers of electrons, it is not clear whether the apparent saturation in $\Delta$ is physical or an artifact of insufficient convergence.
\end{thebibliography}
\end{document}